\documentclass[aps,preprint,preprintnumbers,nofootinbib,showpacs]{revtex4}
\usepackage{graphicx}

\def\beq{\begin{equation}}
\def\eeq{\end{equation}}
\def\beqa{\begin{eqnarray}}
\def\eeqa{\end{eqnarray}}

\def\lsim{\mathrel{\raise.3ex\hbox{$<$\kern-.75em\lower1ex\hbox{$\sim$}}} }
\def\gsim{\mathrel{\raise.3ex\hbox{$>$\kern-.75em\lower1ex\hbox{$\sim$}}} }

\begin{document}
\draft 
\preprint{{\vbox{\hbox{NCU-HEP-k007}\hbox{NSC-NCTS-030217}
\hbox{Feb 2003}
\hbox{rev. May 2003}
}}}
\vspace*{1in}

\title{Can the Two-Higgs-Doublet Model Survive 
the Constraint from the Muon Anomalous Magnetic Moment as Suggested ?\vspace*{.3in} }
\author{Kingman Cheung$^1$ and Otto C. W. Kong$^2$\vspace*{.2in}}
\affiliation
{$^1 \!\!$ National Center for Theoretical Sciences, National Tsing Hua University, Hsinchu, Taiwan.\\ 
$^2 \!\!$ Department of Physics, National Central University, Chung-li, Taiwan 32054\\
Email: {cheung@phys.cts.nthu.edu.tw \& otto@phy.ncu.edu.tw}
\vspace*{.5in}}

\begin{abstract}
Requiring the two-Higgs-doublet model II to accommodate the
3$\sigma$ deviation in the muon anomalous magnetic moment
 imposes specific constraints on the Higgs spectrum. 
We analyze the combination of all the relevant, available, constraints 
on the model parameter space. The use of constraints from
$b\to s\,\gamma$, the precision electroweak measurements of 
$R_b$, and the $\rho$ parameter, together with exclusions from
direct searches at LEP, give extremely severe restrictions
on the model parameters.  That is ``almost enough" to kill the 
model altogether. The exclusion would be even stronger if the
direct searches can be optimized to complement the
other constraints, as will be discussed in details in this work.
\end{abstract}
\pacs{}

\maketitle

\section{Introduction}
Physicists have been delighted by the extraordinary success of
the standard model (SM), while many got frustrated by the
lack of experimental clues for the construction of the theory
beyond.  
The most recent measurement on the muon anomalous magnetic moment 
$a_\mu$ \cite{BNL} revealed a plausible deviation as large as $3\sigma$ 
from the SM prediction, as suggested by the papers in Ref.\cite{amudata}.  
While such a scenario is only a favorable, rather than unquestionably 
established, conclusion from the analyses, it has been taken by 
some physicists as a strong suggestion for physics beyond the SM.
\footnote{
This $3\sigma$ deviation was derived using the $e^+ e^-$ data.  If the $\tau$ 
decay data are used, the deviation would be reduced to about $0.9\sigma$
\cite{amudata}, which, however, has model dependence and thus less reliable.
} 
Since many 
extensions of the SM are capable of giving rise to such a
deviation, theorists, typically, would like to check the constraints 
imposed on the parameter space of a specific model. Such
constraints are particularly interesting, because they are likely to
give information not only on excluded regions but also on the
predictions for where else should other evidences on the model
be expected.  For example, one expects to find a lower bound on the
mass of a new particle playing the role on generating the extra
contribution to  $a_\mu$. The information has especially
strong implications on models that have a small number of
parameters which are already stringently constrained by various precision 
electroweak data. We present here such a case study, illustrating how 
far such a $3\sigma$ deviation can take us.

One of the simplest extensions of the SM is the
two-Higgs-doublet model (2HDM) \cite{hunter}, which adds one Higgs 
doublet in addition to the one required in the SM. A generic 2HDM allows
flavor-changing neutral currents (FCNC), which can be avoided by restricting
the couplings of the doublets, say, by imposing an {\it ad hoc} discrete 
symmetry\cite{glas}. The most popular version, known as model II,
has one Higgs doublet coupled to the down-type quarks (and charged leptons) and 
the second doublet to the up-type quarks.  The physical content of the Higgs sector
(assuming no CP violation) includes a pair of CP-even neutral Higgs bosons 
$H$ and $h$, a CP-odd neutral 
boson $A$, and a pair of charged-Higgs bosons $H^\pm$. The model
fits in well with the criteria mentioned above for the  $a_\mu$
result to have a very strong impact. We focus on the 2HDM II in this paper,
answering the question of to what extent the model can survive a 
requirement of generating the $3\,\sigma$ deviation in $ a_\mu$, as suggested.

The 2HDM II has been extensively studied in literature and tested
experimentally.  One of the most stringent tests is the radiative decay of
$B$ mesons, specifically, the inclusive decay rate of $b\to s\, \gamma$,
which has the least hadronic uncertainties.  In the 2HDM,
the rate of $b\to s\,\gamma$ can be enhanced substantially for large regions
in the parameter space of the mass $m_{\!\scriptscriptstyle H^\pm}$ of
the charged-Higgs boson and 
$\tan\! \beta (=v_{\!\scriptscriptstyle 2}/v_{\!\scriptscriptstyle 1}$,  where 
$v_{\!\scriptscriptstyle 1}$ and $v_{\!\scriptscriptstyle 2}$ are the vacuum 
expectation values of the down-  and up-sector Higgs doublets, respectively).
An earlier analysis has already put a constraint on the charged-Higgs boson 
mass at $m_{\!\scriptscriptstyle H^\pm}> 380$ GeV\cite{cdgg} (see also 
Ref.\cite{nlo,misiak}). Updating the constraint while asking for the model to give 
rise to the $a_\mu$ deviation as suggested already imposes a strong and specific 
mass hierarchy between the pseudoscalar and the charged scalar. Electroweak 
precision data also have strong implications on the model.

The 2HDM can explain the muon anomalous magnetic moment deviation with a 
light pseudoscalar boson $A$ contributing via a 2-loop Barr-Zee-type diagram 
\cite{darwin,ours}.  Our interest here is in updating a previous analysis\cite{ours} 
and extending it to a comprehensive treatment of all the relevant constraints.
The OPAL Collaboration \cite{opal} has recently published an update on their search
for the Higgs bosons within the 2HDM framework.  Since their result is more 
stringent than before, by combining the updated constraints from $a_\mu$ and 
other precision measurements, such as the $\rho$  parameter, $R_b$, and the 
$b\to s\,\gamma$ rate, together with this new OPAL result and a study on Yukawa  
processes from DELPHI \cite{delphi}, we are able to limit the 2HDM to a tiny window 
of parameter space with an, perhaps, uncomfortably large value of $\tan\!\beta$.
All in all, we will piece together a story of the very stringently  constrained 2HDM, 
almost to the extent of killing it altogether. We note that there have been previous
analyses on the 2HDM using the electroweak precision data \cite{maria-ew,chan,maria},
to which we are partially in debt. Our study here can be considered an update, 
with a presentation along a different line.

We are here talking about the intricate interplay of a few stringent constraints on 
the overall parameter space of the model. It is not a simple matter to illustrate
results on the space of a large number (six, as discussed in the
next section) of parameters. In our opinion, the best way to do it 
may depend on how each of the constraints really works. 
Our presentation of the constraints follows what we consider the
most efficient way to appreciate the overall results. It may be not
very conventional, but is considered particularly illustrative. 
Wherever explicit plots are shown, we are typically plotting two-
parameter fits of one or more constraints, based on the usual
$\chi^2$-analysis. We will show $\chi^2<4$ regions, 
corresponding to $2\sigma$ deviation limits, which we consider as
"solution" regions --- where the 2HDM survives the particular
constraints. We will also show regions where the model fits better
than the SM, wherever appropriate. 

The organization is as follows.  In the next section we briefly describe 
the 2HDM (II) and the relevant parameters used in our analysis.
In Sec.~III, we look at the $b\to s\,\gamma$ rate and the $B^0-\overline{B^0}$ 
mixing, which require a heavy charged Higgs boson.
In Sec.~IV,  we discuss the Higgs-sector contributions to $a_\mu$ and
$R_b$, and show the strongly complementary character of the two data
and how that plays off in the 2HDM. We give our fits to the data, with the 
exclusions from DELPHI and OPAL further imposed.   In Sec.~V, we add 
the consideration of the $\rho$ parameter. We conclude in Sec.~VI.

\section{Parameter Space of The Two Higgs Doublet model II }
We take the parameter space of the model as given by a set of six 
Higgs-sector parameters
\[
m_h\;, \;\;
m_{\!\scriptscriptstyle H} \;, \;\;
m_{\!\scriptscriptstyle A} \;, \;\;
m_{\!\scriptscriptstyle H^\pm} \;, \;\;
\tan\!\beta\;,  \;\; \mbox{and} \quad \alpha\;.
\]
Here, the first four are masses of the physical Higgs states. The last one,  
$\alpha$, is the real scalar Higgs mixing angle as defined in 
Ref.\cite{hunter}.  We would like to emphasize that we are
taking this set of six parameters as mutually independent experimental 
parameters. From the theoretical point of view, one has parameters in the 
scalar potential from which the above can be derived. However, without
supersymmetry or anything else to avoid the hierarchy problem, the 
masses for the Higgs states suffer from quadratic divergences. The tree-level 
relations among parameters are modified substantially by loop corrections
which depend on the renormalization approach and the cut-off imposed. To stay 
away from such uncertainties, we do not discuss the scalar potential here, except
noting that there are enough degrees of freedom in the model in general, and 
especially with the loop corrections taken into account, to allow us 
to take the above six parameters as mutually independent.
Here we only consider the 2HDM-II without CP-violation, and are not
interested in couplings among the Higgs states. It is then obvious
that we do not have to consider more than the six parameters. It should
be noted that any substantial CP violation in the Higgs sector that 
may largely invalidate our analysis here is ruled out by the 
electron electric dipole moment constraint.

The four physical masses are direct experimentally measurable 
quantities. The other two parameters, $\tan\!\beta$ and the angle $\alpha$,
come into the game as effective couplings. 
The Yukawa couplings of $h, H$, and $A$ to up- and down-type quarks are
given by, with a common factor of  $-i g m_f/2M_{\!\scriptscriptstyle W}$,
\[
  \begin{tabular}{cccc}
   &  $t \bar t$ &  $b \bar b$ & $\tau^- \tau^+$ \\
$h$: \quad & $\; {\cos\!\alpha/\sin\!\beta}\;\;  $ & $\; \; {- \sin\!\alpha/\cos\!\beta}\;\;  $ & $\; \; {- \sin\!\alpha/\cos\!\beta}\;\;  $\\
$H$: \quad & ${\sin\!\alpha/\sin\!\beta}$ &  $ {\cos\!\alpha/\cos\!\beta}$ &  $ {\cos\!\alpha/\cos\!\beta}$\\
$A$: \quad & $-i \cot\!\beta \,\gamma_5$ & $-i \tan\!\beta  \,\gamma_5$ & $-i \tan\!\beta  \,\gamma_5$ 
  \end{tabular}
\]
while the charged Higgs $H^-$ couples to $t$ and $\bar{b}$  via
\[
\bar b t H^- \,: \;\;\;\; \frac{ig}{2\sqrt{2} M_{\!\scriptscriptstyle W}}\, \left[
  m_t \cot\!\beta \;(1+ \gamma_5) + m_b \tan\!\beta \;(1-\gamma_5) \right ] \;.
\]
From the perspective of our study here, the couplings given above may be 
considered as defining implicitly the two parameters $\tan\!\beta$ and $\alpha$.
Other relevant couplings in our study are those to gauge bosons, 
as given by,
\begin{eqnarray*}
h Z Z \, &:\quad& ig\; M_{\!\scriptscriptstyle Z} {\sin(\beta -\alpha)\over\cos\!\theta_{\!\scriptscriptstyle W} }\; g^{\mu\nu} \nonumber\\
H Z Z \, &:\quad& ig\; M_{\!\scriptscriptstyle Z} {\cos(\beta -\alpha)\over\cos\!\theta_{\!\scriptscriptstyle W}} \; g^{\mu\nu} \nonumber\\
h A Z \, &:\quad& g\;  {\cos(\beta -\alpha)\over 2\cos\!\theta_{\!\scriptscriptstyle W}} \; (p-p')^\mu \nonumber\\
H A Z \, &:\quad& -g \;{\sin(\beta -\alpha)\over 2\cos\!\theta_{\!\scriptscriptstyle W}} \; (p-p')^\mu \nonumber\\
H^+ H^- Z \, &:\quad& -ig\; {\cos\! 2\theta_{\!\scriptscriptstyle W}\over 2\cos\!\theta_{\!\scriptscriptstyle W}} \; (p-p')^\mu \;,
\end{eqnarray*}
where $p(h,H,H^+)$ and $p'(A,H^-)$ are the 4-momenta going into the vertex.

\section{Requirement of a heavy charged Higgs Boson}
It has been well appreciated that $B$ physics bars the 2HDM from admitting
a relatively light charged-Higgs state. We first review the constraints here.
The first important constraint comes 
from the inclusive $B\to X_s\gamma$ result, and the second one
comes from the $B^0-\overline{B^0}$ mixing. 
The essential point here is that without a direct source
of FCNC, the charged Higgs mediates the only significant contributions
to flavor-changing processes, in addition to the $W^\pm$-mediated SM  
process. The experimental data then allows us to bound the charged
Higgs mass independent of the other Higgs states.

The detail description of the effective Hamiltonian approach can be found in
Refs. \cite{buras,wise}.  Here we present the highlights that are relevant
to our discussions.  The effective Hamiltonian for $B\to X_s\gamma$ at a
factorization scale of order $O(m_b)$ is given by
\begin{equation}
\label{eff}
{\cal H}_{\rm eff} = - \frac{G_{\!\scriptscriptstyle F}}{\sqrt{2}} V^*_{ts} V_{tb} \biggr [
\sum_{i=1}^6 \; C_i(\mu) Q_i(\mu) + C_{7\gamma}(\mu) Q_{7\gamma}(\mu) +
C_{8G}(\mu) Q_{8G}(\mu) \biggr ] \;.
\end{equation}
The operators $Q_i$ can be found in Ref.\cite{buras}, of which the $Q_1$ and
$Q_2$ are the current-current operators and $Q_3-Q_6$ are QCD penguin
operators.  $Q_{7\gamma}$ and $Q_{8G}$ are, respectively, the magnetic
penguin operators specific for $b\to s\,\gamma$ and $b\to s\, g$.  Here we also
neglect the mass of the external strange quark compared to the external
bottom-quark mass. 
There have been more recent analyses \cite{misiak,gambino,urban}
on $b\to s \,\gamma$ involving the NLO
and other corrections, but the LO treatment here is sufficient for our 
purpose to put a lower bound on the charged Higgs mass $m_{H^\pm}$, which
is then used in the central part of our analysis.

The decay rate of $B\to X_s \gamma$ normalized to the
experimental semileptonic decay rate is given by
\begin{equation}
\label{cal}
\frac{\Gamma(B\to X_s\gamma)}{\Gamma(B\to X_c e \bar \nu_e)} =
\frac{|V^*_{ts} V_{tb}|^2}{|V_{cb}|^2} \, \frac{6\,\alpha_{\rm em}}
 {\pi f(m_c/m_b)} |C_{7\gamma}(m_b)|^2 \;,
\end{equation}
where $f(z) = 1-8z^2 +8z^6 - z^8 -24z^4 \ln z$.
The Wilson coefficient $C_{7\gamma}(m_b)$ is given by
\begin{equation}
\label{c7}
C_{7\gamma}(\mu) = \eta^{\frac{16}{23}} C_{7\gamma}(M_{\!\scriptscriptstyle W}) + \hbox{$8\over3$}
\left(\eta^{\frac{14}{23}} - \eta^{\frac{16}{23}} \right ) C_{8G}(M_{\!\scriptscriptstyle W})
+ C_2(M_{\!\scriptscriptstyle W}) \sum_{i=1}^8 h_i \eta^{a_i} \;,
\end{equation}
where $\eta=\alpha_s(M_{\!\scriptscriptstyle W})/\alpha_s(\mu)$.  The $a_i$'s and $h_i$'s  
can be found in Ref.~\cite{buras}.
The coefficients $C_i(M_{\!\scriptscriptstyle W})$ at the leading order in 2HDM II 
are given by
\begin{eqnarray}
C_j(M_{\!\scriptscriptstyle W}) &=& 0 \qquad\qquad (j=1,3,4,5,6) \;, \label {c6mw}\\
C_2(M_{\!\scriptscriptstyle W}) &=& 1 \;, \\
C_{7\gamma}(M_{\!\scriptscriptstyle W}) &=& - \frac{A(x_t)}{2} - \frac{A(y_t)}{6} \cot^2\!\beta  - B(y_t) \;, \\
C_{8G} (M_{\!\scriptscriptstyle W}) &=& -\frac{D(x_t)}{2} - \frac{D(y_t)}{6} \cot^2\!\beta      - E(y_t) \;,
\end{eqnarray}
where $x_t = m_t^2/M_{\!\scriptscriptstyle W}^2$, and 
$y_t = m_t^2 / m_{\!\scriptscriptstyle H^\pm}^2$.
The Inami-Lim functions\cite{japan} are given by
\begin{eqnarray}
A(x) &=& x \biggr [ \frac{8x^2 +5x -7}{12(x-1)^3} - \frac{(3x^2 -2x)\ln x}
                           {2(x-1)^4} \biggr ] \;, \\
B(y)&=&y\biggr[\frac{5y-3}{12 (y-1)^2} - \frac{(3y-2) \ln y}{6(y-1)^3}\biggr ] \;, \\
D(x)&=& x\biggr[ \frac{x^2 - 5x -2}{4(x-1)^3} + \frac{3x\ln x}{2(x-1)^4}\biggr] \;, \\
E(y)&=& y\biggr[\frac{y-3}{4(y-1)^2} + \frac{\ln y}{2(y-1)^3} \biggr ] \;.
\end{eqnarray}

The most recent experimental data on $b\to s \,\gamma$ rate has been reported
\cite{ichep-b}, giving
\[
B(b\to s \,\gamma) |_{\rm exp} = 3.88 \pm 0.36 ({\rm stat}) \pm 0.37 ({\rm sys})
^{+0.43}_{-0.28} ({\rm theory}) \;.
\]
The most updated SM prediction is \cite{neubert}
\[
B(b\to s \,\gamma) |_{\rm \small SM} = (3.64 \pm 0.31 ) \times 10^{-4} \;,
\]
which agrees very well the data.  
Both the experimental data and the SM prediction have been extrapolated to 
the total branching ratio.
Therefore, there is only a little room for new 
physics contributions.   The constraint on new physics contribution is, explicitly, 
\begin{equation}
\Delta B (b\to s\,\gamma) \equiv 
B(b\to s\, \gamma) |_{\rm exp} - B(b\to s \,\gamma) |_{\rm SM}
= (0.24  ^{+0.67}_{-0.59} ) \times 10^{-4} \;,
\end{equation}
where we have added the various errors of the experimental data in quadrature.
(Note that the theory error quoted in the experimental data is larger than
the one quoted by the SM prediction.  We take the more conservative value.)

The quantity that parameterizes the $B^0-\overline{B^0}$ mixing is
\begin{equation}
x_d \equiv \frac{\Delta m_{\!\scriptscriptstyle B}}{\Gamma_{\!\scriptscriptstyle B}} 
= \frac{G_{\!\scriptscriptstyle F}^2}{6\pi^2} |V_{td}^*|^2
|V_{tb}|^2 f_{\!\scriptscriptstyle B}^2 \, B_{\!\scriptscriptstyle B} \, m_{\!\scriptscriptstyle B}
\eta_{\scriptscriptstyle B} \tau_{\!\scriptscriptstyle B} \, M_{\!\scriptscriptstyle W}^2 
\left(I_{\!\scriptscriptstyle W\!W} + I_{\!\scriptscriptstyle W\!H}+I_{\!\scriptscriptstyle H\!H} \right) \;,
\end{equation}
where\cite{abbott}
\begin{eqnarray}
I_{\!\scriptscriptstyle W\!W} &=& \frac{x}{4} \biggr[ 1 + \frac{3-9x}{(x-1)^2} +
 \frac{6x^2 \log x}{(x-1)^3} \biggr] \;, \nonumber \\
I_{\!\scriptscriptstyle W\!H} &=& xy \cot^2\!\beta \biggr[ \frac{(4z-1)\log y}{2(1-y)^2(1-z)}
 -\frac{3\log x}{2(1-x)^2(1-z)} +\frac{x-4}{2(1-x)(1-y)} \biggr] \;, \nonumber \\
I_{\!\scriptscriptstyle H\!H} &=& \frac{xy \cot^4\!\beta }{4} \biggr[ \frac{1+y}{(1-y)^2} +
  \frac{2y\log y}{(1-y)^3} \biggr]\nonumber  \;,
\end{eqnarray}
with $x=m_t^2/M_{\!\scriptscriptstyle W}^2$, $y=m_t^2/m_{\!\scriptscriptstyle H^\pm}^2$, $z=M_{\!\scriptscriptstyle W}^2/m_{\!\scriptscriptstyle H^\pm}^2$, and
the running top mass $m_t=m_t(m_t)=166\pm 5$ GeV.
The experimental value is \cite{pdg2002}
\begin{equation}
x_d=0.755\pm 0.015 \;.
\end{equation}
We use the following input parameters \cite{pdg2002}: 
$|V_{tb} V_{td}^*|=0.0079\pm 0.0015$, 
$f_{\!\scriptscriptstyle B}^2 B_{\!\scriptscriptstyle B}= (198\pm30\;{\rm GeV})^2(1.30\pm0.12)$, 
$m_{\!\scriptscriptstyle B}=5279.3\pm0.7$ MeV, $\eta_{\scriptscriptstyle B}=0.55$, and
$\tau_{\!\scriptscriptstyle B}=1.542\pm0.016$ ps.
Note that 
the value of $|V_{tb} V_{td}^*|$ is in fact determined by the measurement of
$x_d$.  Now we can use the data to constrain the new contribution from
the charged-Higgs boson.

\begin{figure}[th!]
\includegraphics[width=6in]{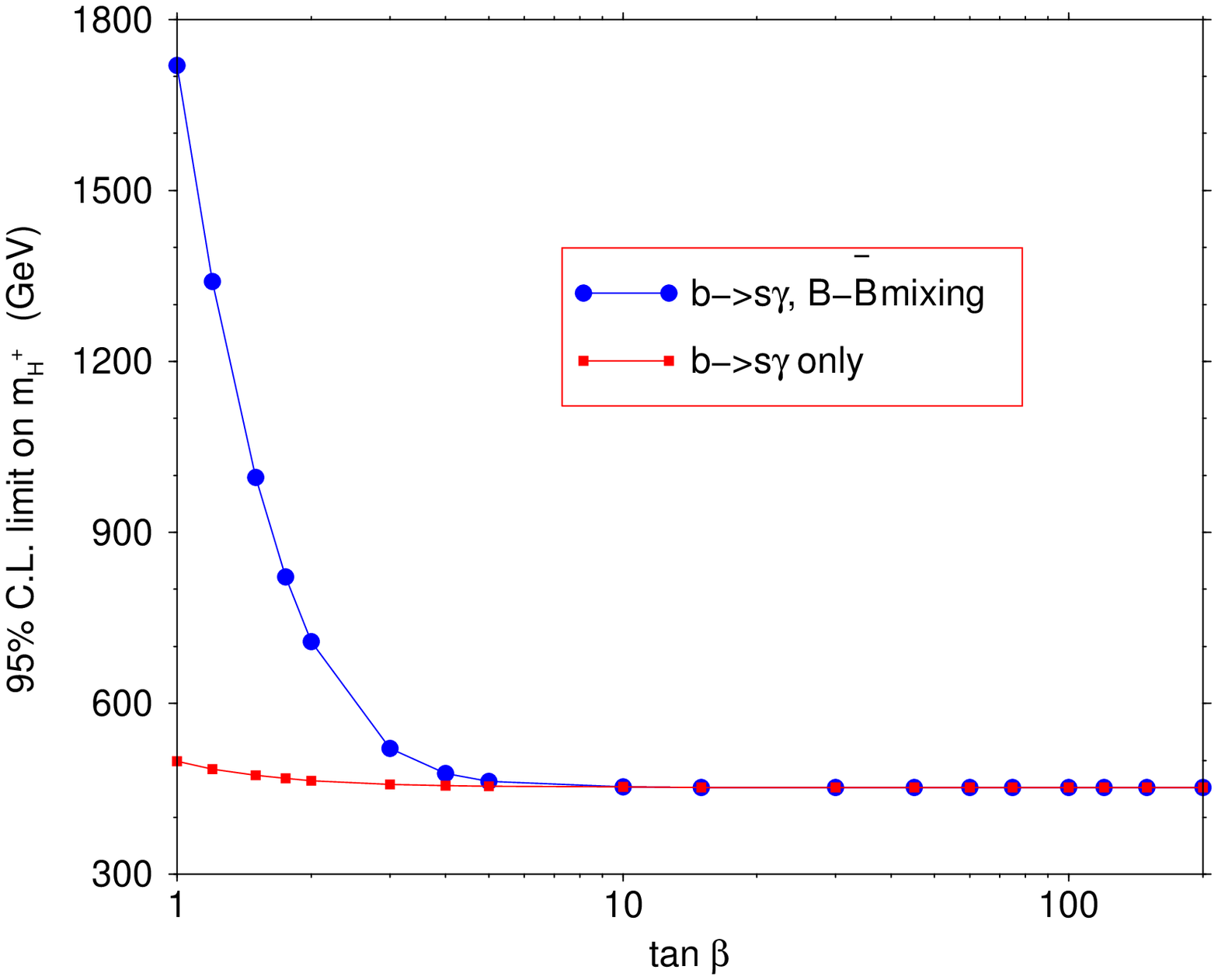}
\caption{\small
95\% C.L. lower limit on the charged Higgs mass vs $\tan\!\beta$
due to the constraints on $b\to s\,\gamma$ and/or $B-\overline{B}$ mixing.
\label{fig_1}
} 
\end{figure}

The two constraints discussed above are quite stringent, giving a lower bound 
on $m_{\!\scriptscriptstyle H^\pm}$ close to $500\,\mbox{GeV}$ at 95\% C.L.
for intermediate and large values of $\tan\!\beta$.
The result is illustrated in Fig.~\ref{fig_1}. 
The heavy charged Higgs mass means that
the state pretty much decouples, playing a little role in the contributions of
the 2HDM to quantities like $a_\mu$ and $R_b$, which we
turn to in the next section.

\section{\boldmath\protect $\mbox{$a$}_\mu$ Vs $R_{b}$}
The most recent data on the $a_\mu$ indicates \cite{amudata} (see also
the footnote \# 1)
\begin{equation}
\label{33}
\Delta a_\mu \equiv a_\mu^{\rm exp} - a_\mu^{\rm SM} = 
(33.9 \pm 11.2) \times 10^{-10} \;.
\end{equation}
The result shows a $3\sigma$ deviation to be explained by new physics.
Adopting the view that the $a_\mu$ problem is real and demands new physics 
contributions, we will see that it has a strong and definite implication on 
the Higgs spectrum of the 2HDM. In fact, we had performed an analysis 
\cite{ours}
along the line for the earlier data. The major point is that a light 
pseudoscalar, together
with a large $\tan\!\beta$ value, is required to explain the
positive $\Delta a_\mu$ contribution via a two-loop Barr-Zee diagram \cite{BZ}.
A real scalar
contributes in the negative direction. To avoid a cancellation, the real scalar
mass has to be heavy. We will show here that such a mass splitting is 
strongly disfavored by the allowed contribution to $R_b$, for 
which the experimental data agrees well with the SM prediction. The two 
constraints are hence strongly complementary.
   
It has been emphasized in Ref.\cite{darwin,ours} that for the Higgs boson mass 
larger than about $3\,\mbox{GeV}$, the dominant Higgs 
contributions to  $a_\mu$ actually come from the two-loop Barr-Zee diagram 
with a heavy fermion ($f$) running in the upper loop. A $m_f^2/m_\mu^2$ factor 
could easily overcome the $\alpha/4\pi$ loop factor. In our calculation here,
we include all one-loop contributions and all two-loop Barr-Zee-type
contributions
with an internal photon and a third-family fermion running in the loop. The
latter diagrams with the bottom and tau loops
are strongly enhanced by $\tan\!\beta$.  If the internal
photon was replaced by a $W^\pm$ or a $Z^0$, the contributions will be much
suppressed (see \cite{edm}, for examples). The $W^\pm$ case is
in particular strongly suppressed, partly as a result of the fact that
the Higgs boson has to be $H^\pm$, the mass of which we have shown above 
to be heavier than 500 GeV. The only other important diagrams of
the Barr-Zee type are the SM diagrams, such as the one with
the $W^\pm$ replacing the heavy fermion. We neglect the small
``extra" contributions from such diagrams\cite{wy} because of the
small difference between the Higgs boson mass used here and that used in the  
Ref. \cite{M}.

Explicitly, we first write the fermion
couplings of a neutral Higgs mass eigenstate $\phi^0$ as
\begin{equation}
{\cal L}^{ \bar{f}\!\phi^0\! f}
=  - \lambda_{f} \, \frac{m_{\! \scriptscriptstyle f}}{v}
\bar{f}\phi^0 f +
i \gamma_{\scriptscriptstyle 5}\, A_{f}  \,
\frac{m_{\! \scriptscriptstyle f}}{v}
\bar{f} \phi^0 f \; ,
\end{equation}
where $\lambda_{f}\,
\frac{m_{\! \scriptscriptstyle f}}{v}$ and
$A_{f}\, \frac{m_{\! \scriptscriptstyle f}}{v}$ are
the effective scalar and pseudoscalar couplings explicitly
given in Sec. II, and $v=246\,\mbox{GeV}$.  
The two-loop photon Barr-Zee diagram contribution from $\phi^0$, with a
heavy fermion $f$ running in the second loop, is given by
\begin{equation} \label{s}
\Delta a_\mu^{\phi} =
\frac{N_{\!c}^f \, \alpha_{\mbox{\tiny em}}}{4\pi^3 \,v^2} 
{m_\mu^2}\;
{\cal Q}_f^2 \left[ A_\mu \, A_f \,
g\!\!\left( \frac{m_f^2}{m_{\phi}^2} \right)
- \lambda_\mu \, \lambda_f \,
f\!\!\left( \frac{m_f^2}{m_\phi^2} \right) \right]\; ,
\end{equation}
where
\beqa
f(z)={1\over 2} z \int^1_0 \! dx \; 
\frac{1-2x(1-x)}{x(1-x)-z} \ln\frac{x(1-x)}{z} \; ,
\nonumber \\
g(z)={1\over 2} z \int^1_0 \!  dx \; \frac{1}{x(1-x)-z} \ln\frac{x(1-x)}{z} \;;
\eeqa
$N_{\!c}^f$ represents the number of color degrees of freedom in $f$, and
${\cal Q}_f$ its electric charge.  Here, we have three scalars and no CP
violating mixing is assumed. The real scalars
$h$ and $H$ give negative contributions only (from the second part) while
a pseudoscalar $A$ gives a positive contribution only (from the first part).
The diagrams with the $b$ and $\tau$ loops 
are $\tan\!\beta$ enhanced.

\begin{figure}[th!]
\includegraphics[width=6in]{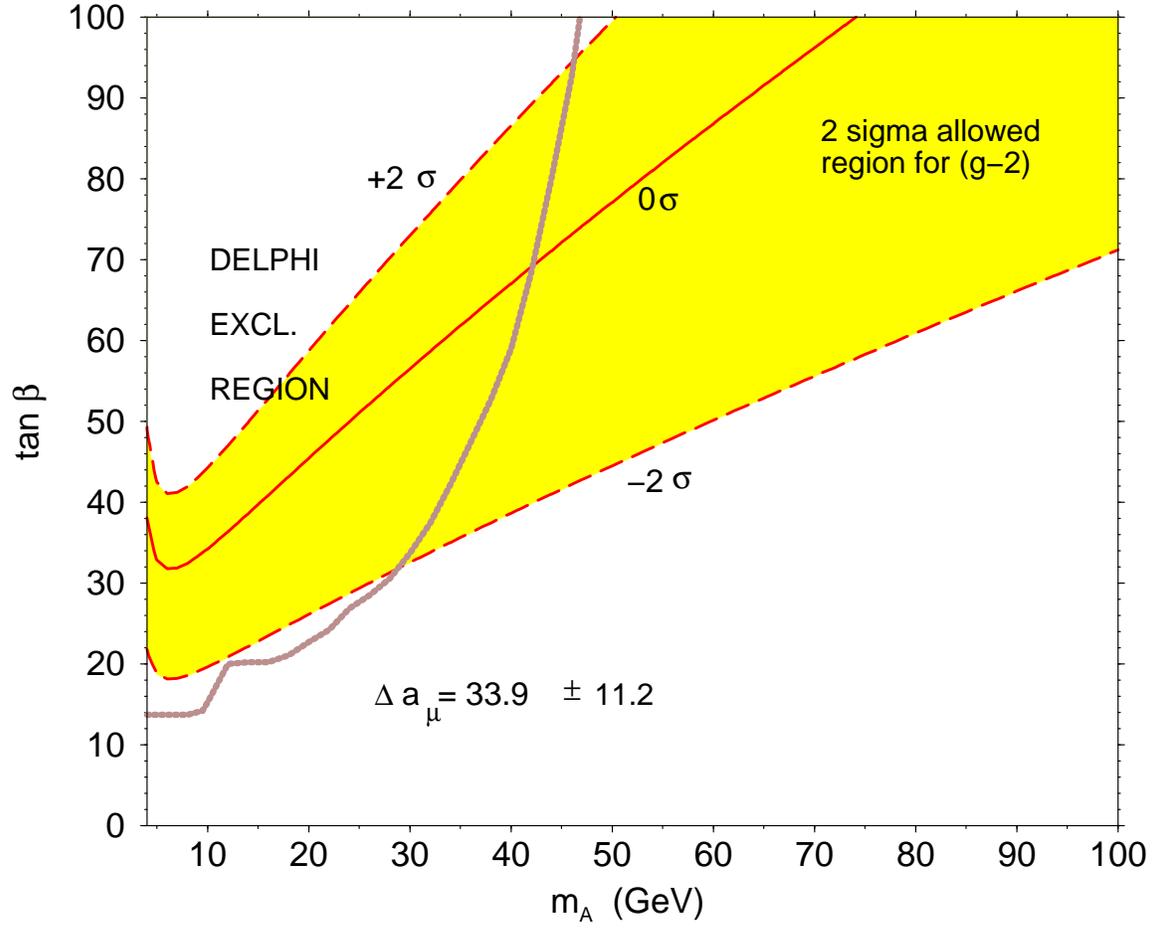}
\caption{\small The $2\sigma$ allowed region in the 
$(m_{\!\scriptscriptstyle A},\; \tan\!\beta)$ 
plane due to the $a_\mu$ data.  Here only the pseudoscalar contribution is 
considered. The DELPHI excluded region is represented by the thick line.} 
\label{fig_2}
\end{figure}

We plot in Fig.~\ref{fig_2}  the $2\sigma$ range of solution to $a_\mu$ 
on the plane of the pseudoscalar mass $m_{\!\scriptscriptstyle A}$ versus 
$\tan\!\beta$, considering only the pseudoscalar contribution. Note that 
while the region below the solution band is excluded, the solution in 
the region above the band  may be admissible when the $a_\mu$ is 
compensated by some negative contributions from the real scalar(s). 
We also superimpose on the plot the excluded region from the DELPHI study 
on Higgs Yukawa processes in the $4b$ and $2b2\tau$ 
final states \cite{delphi}.  We can see that one obtains
lower bounds on $m_{\!\scriptscriptstyle A}$ and $\tan\!\beta$
as $26\,\mbox{GeV}$ and $30$ respectively.
\footnote{The plot is similar to the one given in Ref.~\cite{maria},
in which some more constraints are superimposed. We include here
only the important ones. Note that the Tevatron exclusion region
claimed in the paper is not used here. The exclusion result
was from Ref.\cite{roco}, which is an analysis based on the minimal
supersymmetric standard model. The result should not be 
directly applicable to the present case of 2HDM. We do not find
a similar study on the Tevatron data based on the 2HDM.}
We had given in Ref.~\cite{ours} plots of the solution regions
on the $m_h$-$m_{\!\scriptscriptstyle A}$ plane 
for the old $a_\mu$ data with specific
values of  $\tan\!\beta$ and the scalar mixing angle $\alpha$.
We will present similar results here with, however, 
the complementary $R_b$ constraint included. We will see that
not much area of the parameter space can survive the combination
of both $a_\mu$ and $R_b$ data.

The current $R_b$ measurement is given by \cite{lep-ew}
\[
R_b^{\rm exp} = 0.21646 \pm 0.00065 \;.
\]
With $R_b^{\rm SM} = 0.215768$, we have
\begin{equation}
\Delta R_b \equiv R_b^{\rm exp} - R_b^{\rm SM} = 0.000692 \pm 0.00065 \;.
\end{equation}
The $\Delta R_b$ contributions in the 2HDM are given, for example,
by formulas in Ref.\cite{denner}. The charged Higgs contribution is always negative.  
On the other hand, there is a window of parameter space for the neutral Higgs  
contributions to be positive.  From our discussions above, we need a light
pseudoscalar and have to live with a heavy charged Higgs. We are therefore
more interested in the neutral Higgs contributions. 

\begin{figure}[th!]
\includegraphics[width=3.2in]{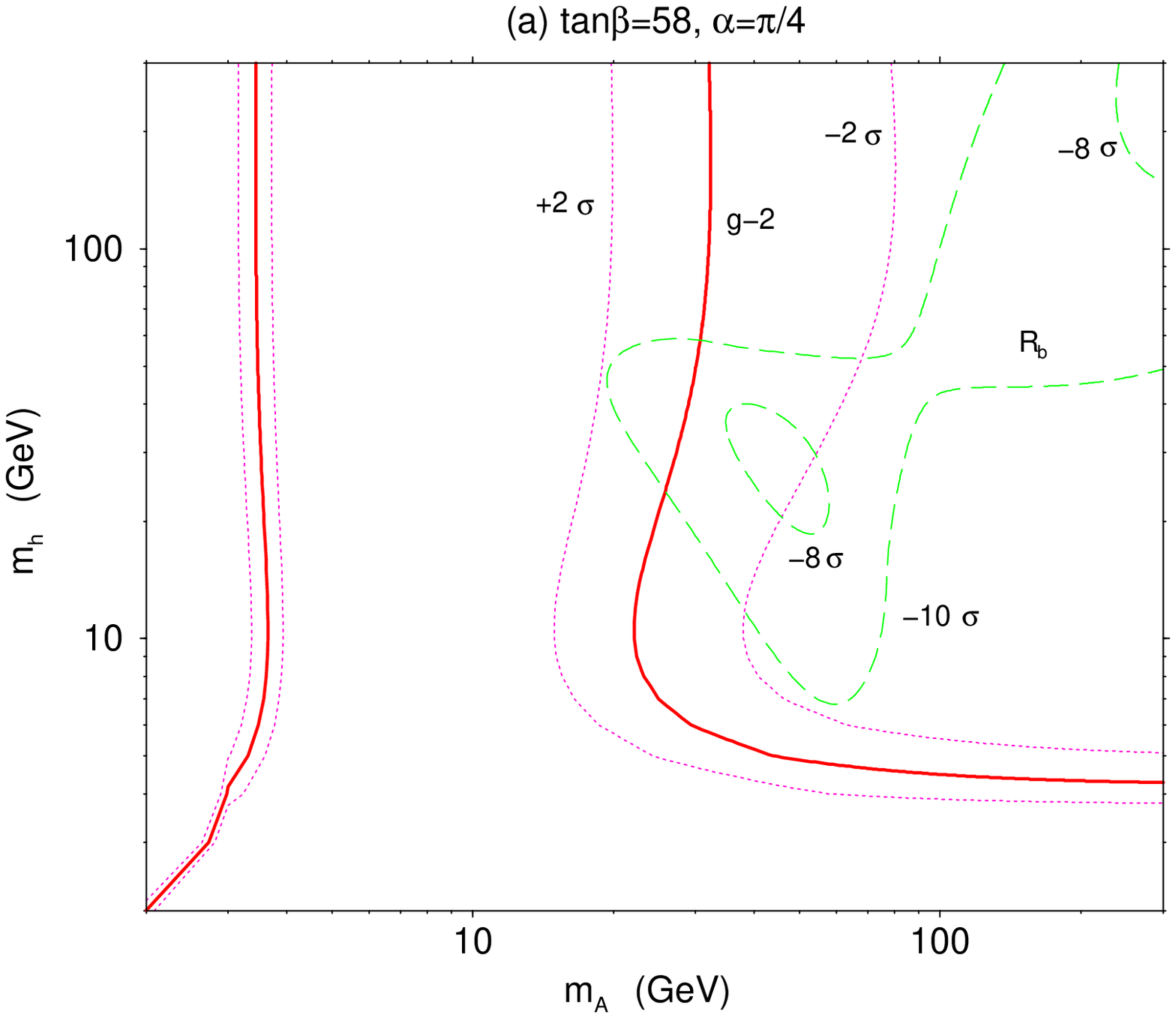}
\includegraphics[width=3.2in]{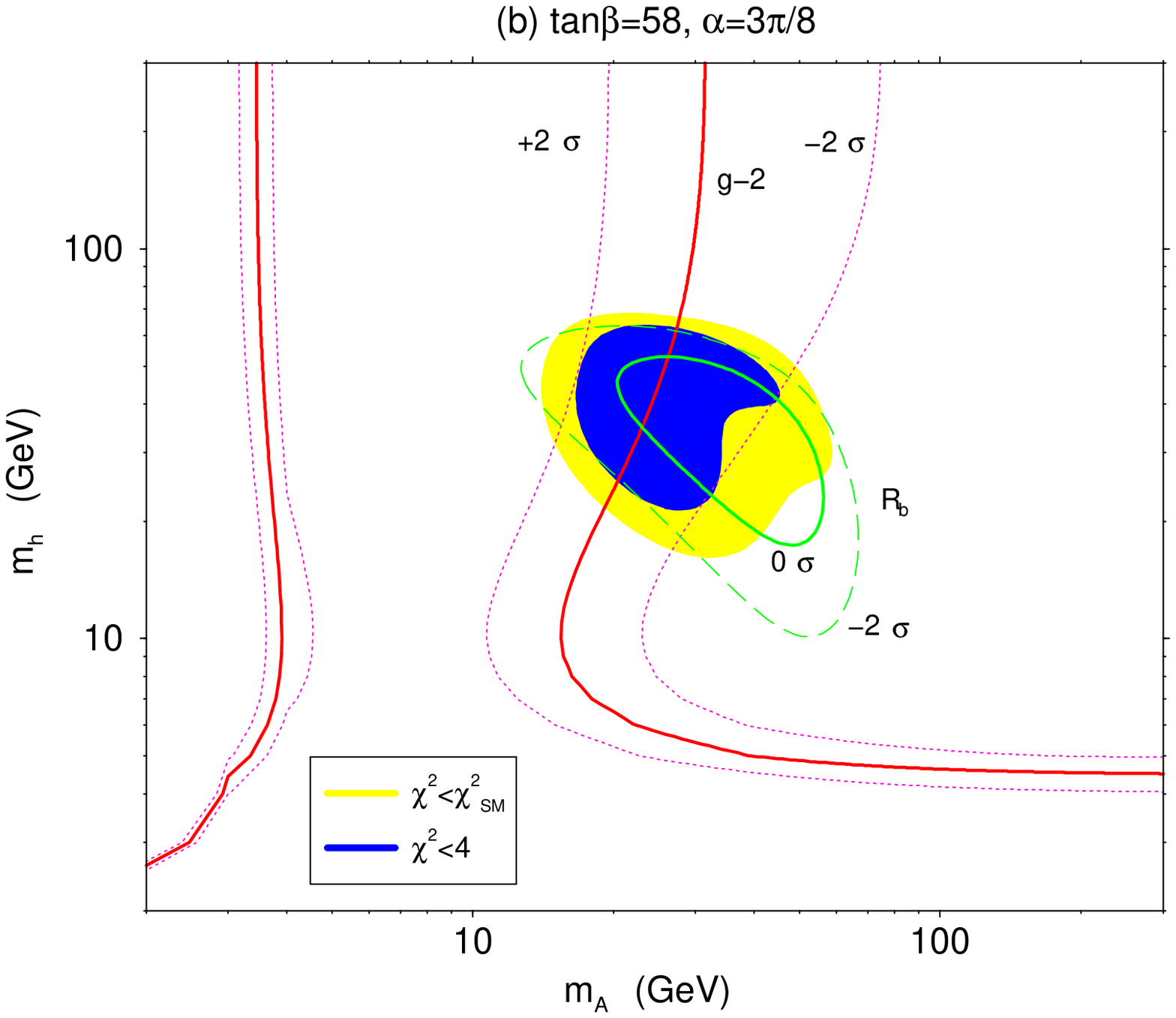}

\includegraphics[width=3.2in]{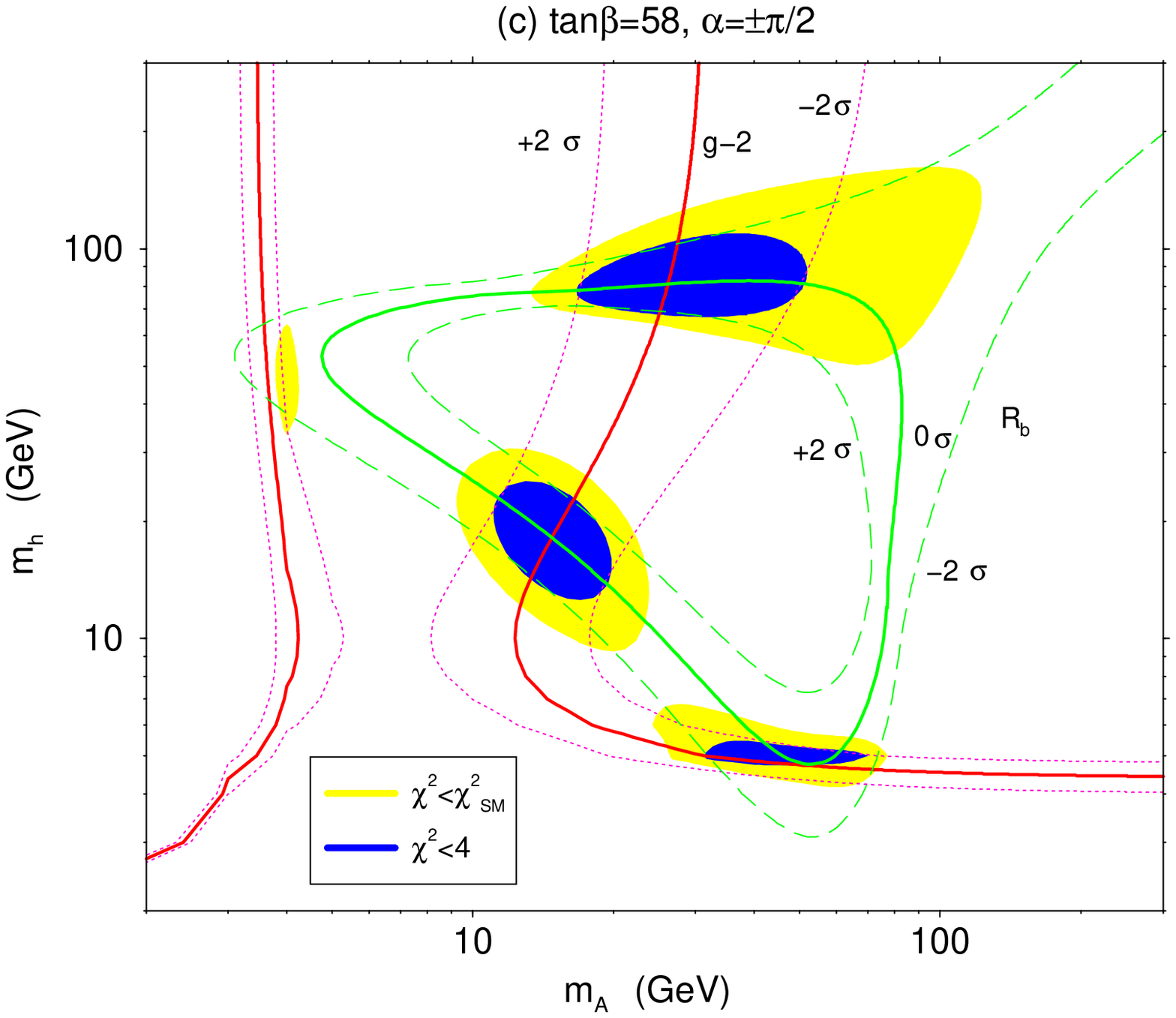}
\includegraphics[width=3.2in]{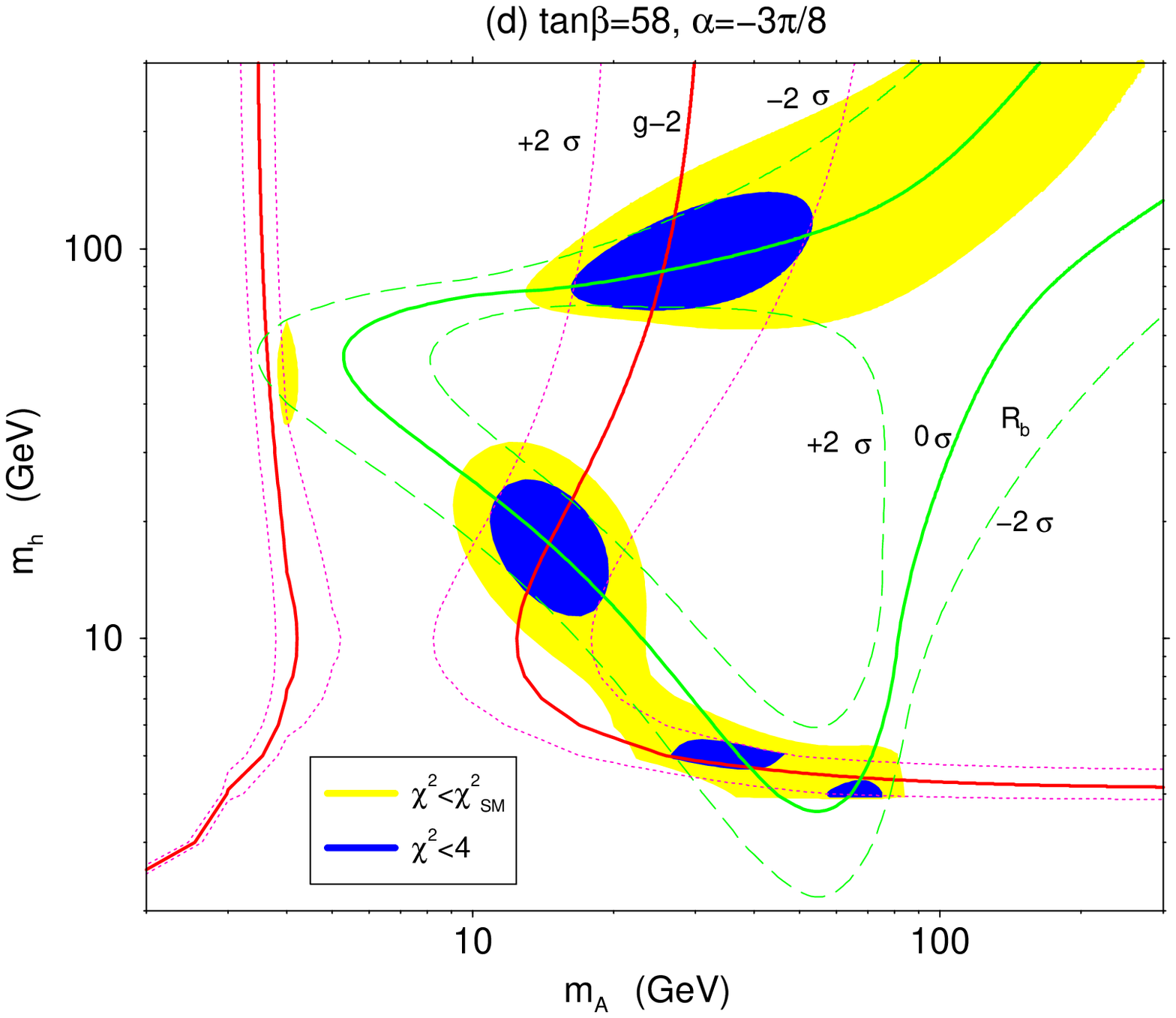}

\includegraphics[width=3.2in]{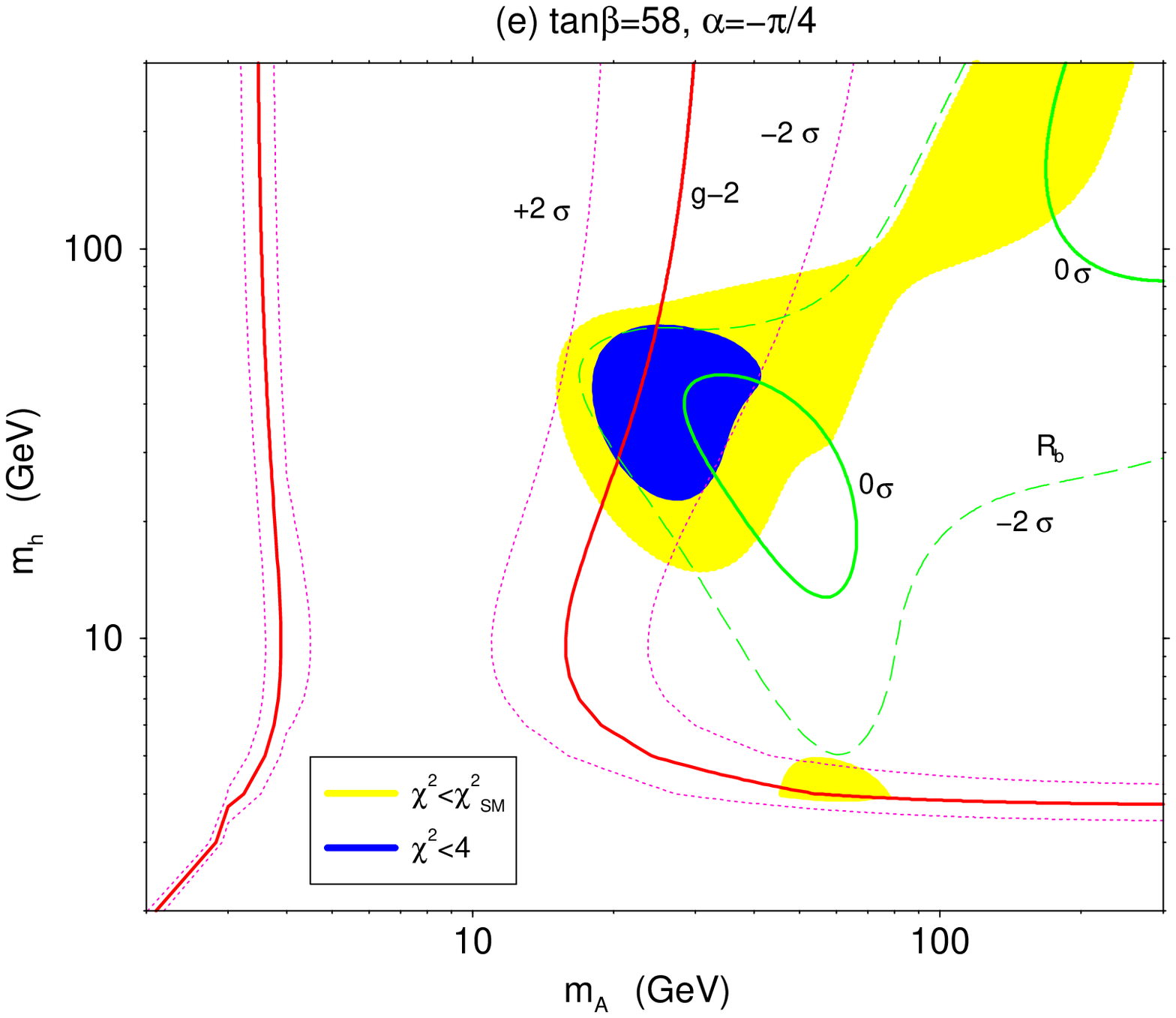}
\includegraphics[width=3.2in]{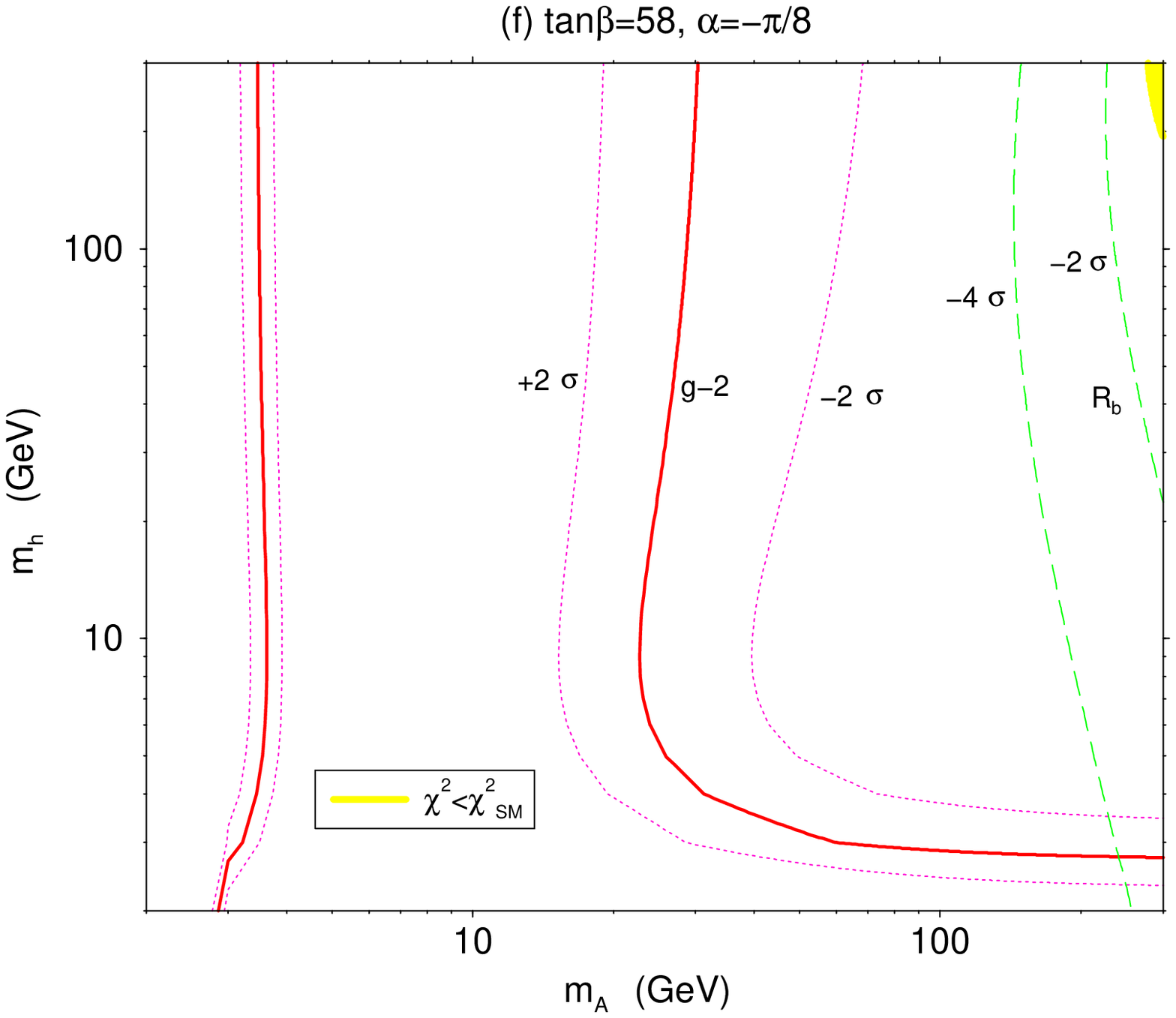}
\caption{\small
The $2\sigma$ allowed regions in the $(m_{\!\scriptscriptstyle A},\;m_h)$ plane due to the
constraints of $a_\mu$ and $R_b$ for $\tan\!\beta=58$.  
The smaller dark region is where
the total $\chi^2$ is less than 4 while the lighter region is where the
total $\chi^2$ is less than the $\chi^2({\rm SM})=10.3$.
} 
\label{fig_3}
\end{figure}

Let us first focus on the contributions from the lighter Higgs bosons, the 
pseudoscalar $A$ and the real scalar $h$, pushing the other scalar $H$ to 
the heavy-mass
limit together with $H^\pm$.  The scenario will actually be well justified by
our discussion on the constraint of the $\rho$ parameter in the next section. 
The smallness of $\Delta R_b$ contributions admitted here generally 
disfavors a large mass splitting between 
$m_h$ and $m_{\!\scriptscriptstyle A}$.
This is in contrast to the requirement of a positive contribution to $a_\mu$.
Since the SM result  (with $M_{H_{SM}}$ at $115\,\mbox{GeV}$) now represents 
a $3\sigma$ deviation in $a_\mu$, better fits to the combined $a_\mu$-$R_b$
data are possible from the 2HDM. We illustrate some such fits in 
Fig.~\ref{fig_3}.
In the figure, we take the case of $\tan\!\beta=58$ and check various
values of the Higgs mixing angle $\alpha$.  Here, and in the discussion
below unless specifically stated otherwise, we stick to 
$m_{\!\scriptscriptstyle H^\pm}=500\,\mbox{GeV}$ and 
$m_{\!\scriptscriptstyle H}=1\,\mbox{TeV}$.  The exact value
of $m_{\!\scriptscriptstyle H}$ does not matter at all here,
one, however,  should note that
the charged Higgs boson still gives a contribution of $-2.32\times10^{-4}$
to $R_b$, which is about ${1\over 3}\sigma$ in strength.  Bearing
this in mind, it is easy to estimate from our plots the slight shift
in each of the admissible region as 
$m_{\!\scriptscriptstyle H^\pm}$ is being pushed towards 
the decoupling limit. 

Each of the plots in Fig.~\ref{fig_3} gives $\pm 2\sigma$ limits for $a_\mu$
and $R_b$ fits, with darker shaded regions indicating the solution 
of interest defined by a total $\chi^2$ of $4$ or less. 
Also marked in the plots are regions with a total $\chi^2$ 
less than the SM value of  $10.3$ (the sum of $a_\mu$ and $R_b$).
The purpose of showing the area with a total $\chi^2 < \chi^2({\rm SM})$
is to indicate the region of parameter space that can fit the $a_\mu$ and
$R_b$ better than the SM, other than the decoupling limits.  From now on,
we concentrate on the dark area of a total $\chi^2 <4$ as a valid solution to 
the $a_\mu$ and $R_b$ data. We can see that there are no solutions for 
$-{\pi\over 8}<\alpha<{\pi\over 4}$. While a larger magnitude of the $\alpha$ 
looks more favorable, the best solution, with higher Higgs masses 
($m_h$ especially), stay close to $|\sin\!\alpha|=1$,
 inclining more towards
the negative sign. The most favorable range is around
$-{\pi\over 2}<\alpha<-{3\pi\over 8}$  [{\it cf.} plots (c) and (d)].
The plots in Fig.~\ref{fig_3} illustrate well the trend of the changes in the 
$a_\mu$-$R_b$ solution regions with variations in $\alpha$, which is quite 
generic for $\tan\!\beta$ around and larger than 50.
A higher $m_h$ solution is preferred as the light mass solutions are
easily killed by searches at LEP, including the DELPHI exclusion used
in Fig.~\ref{fig_2} and a particularly focused 2HDM analysis from 
OPAL\cite{opal}, to be discussed below. In fact, as we will illustrate
below, if any part of the solutions to the $a_\mu$-$R_b$ fits survive 
the exclusions from LEP, it is more or less the part with a high enough 
$m_h$ value.
\footnote{In these plots, we used the standard $R_b$ formulas for 2HDM
as available in the reference \cite{denner} quoted. The formulas have not taken
into consideration the possible tree-level decays of the $Z^0$ boson
into a pair of Higgs states. Such very light Higgs regions are typically
not of interest to theorists, as they would be easily excluded by
direct searches. By the same token, we use the formulas as they are and
show the $a_\mu$-$R_b$ fits without bothering about the tree-level
modifications of $R_b$ from such Higgs channels, leaving such
fake solutions to be taken care of by the exclusions from LEP searches
to be discussed below. We thank A.~Sanda for alerting us to explicitly
address the issue here.} 

\begin{figure}[th!]
\includegraphics[width=3.2in]{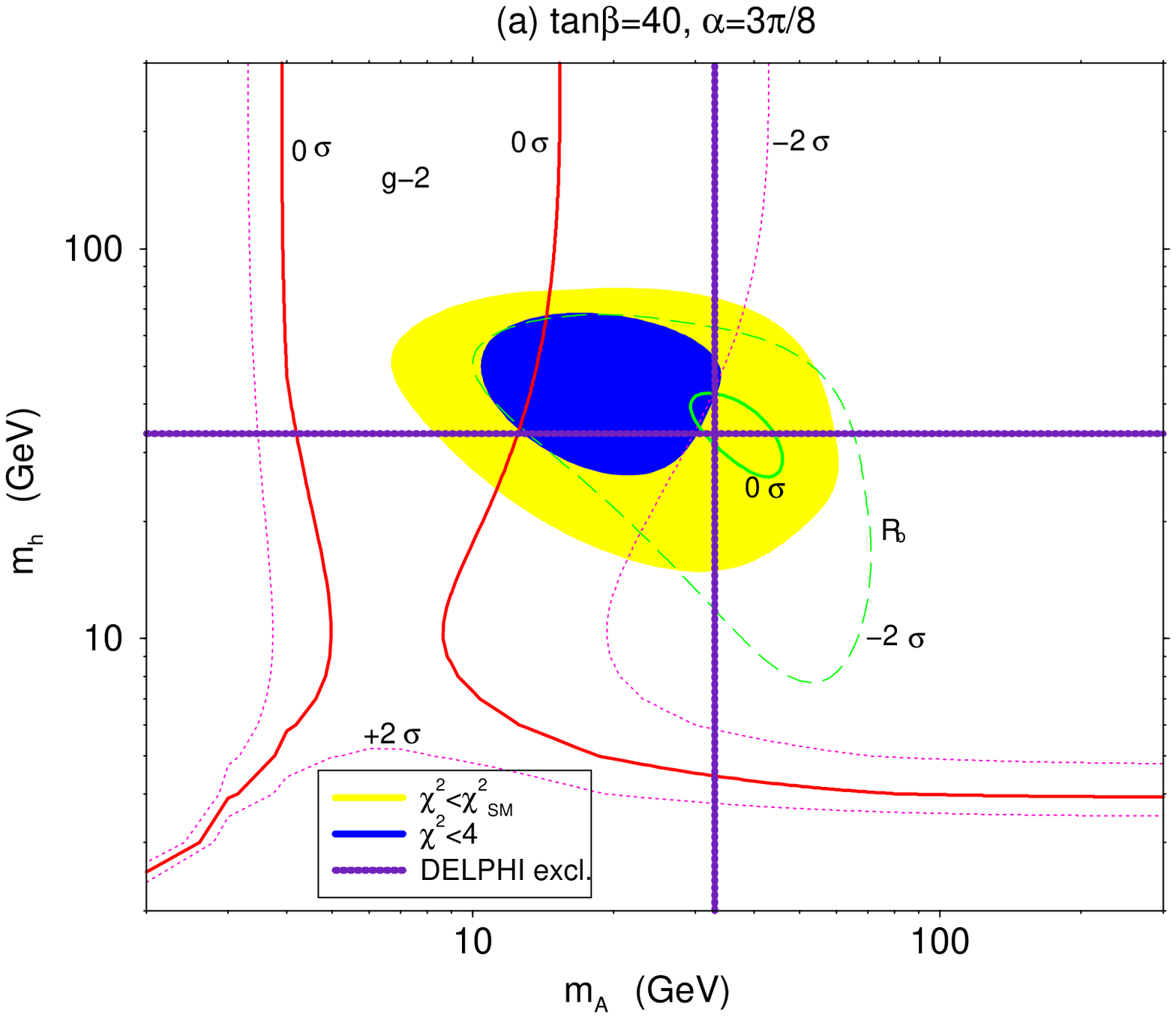}
\includegraphics[width=3.2in]{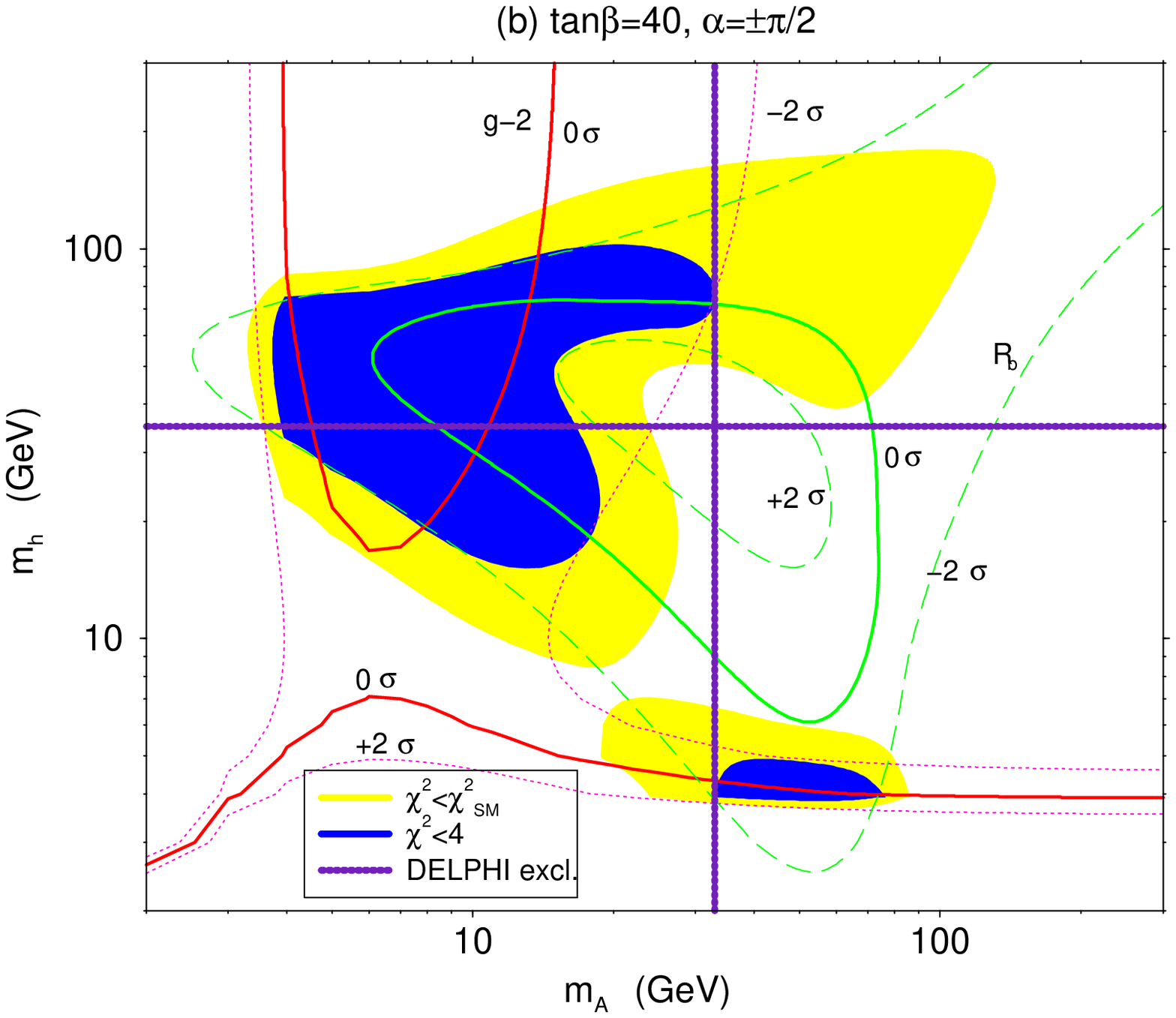}

\vspace{0.2in}

\includegraphics[width=3.2in]{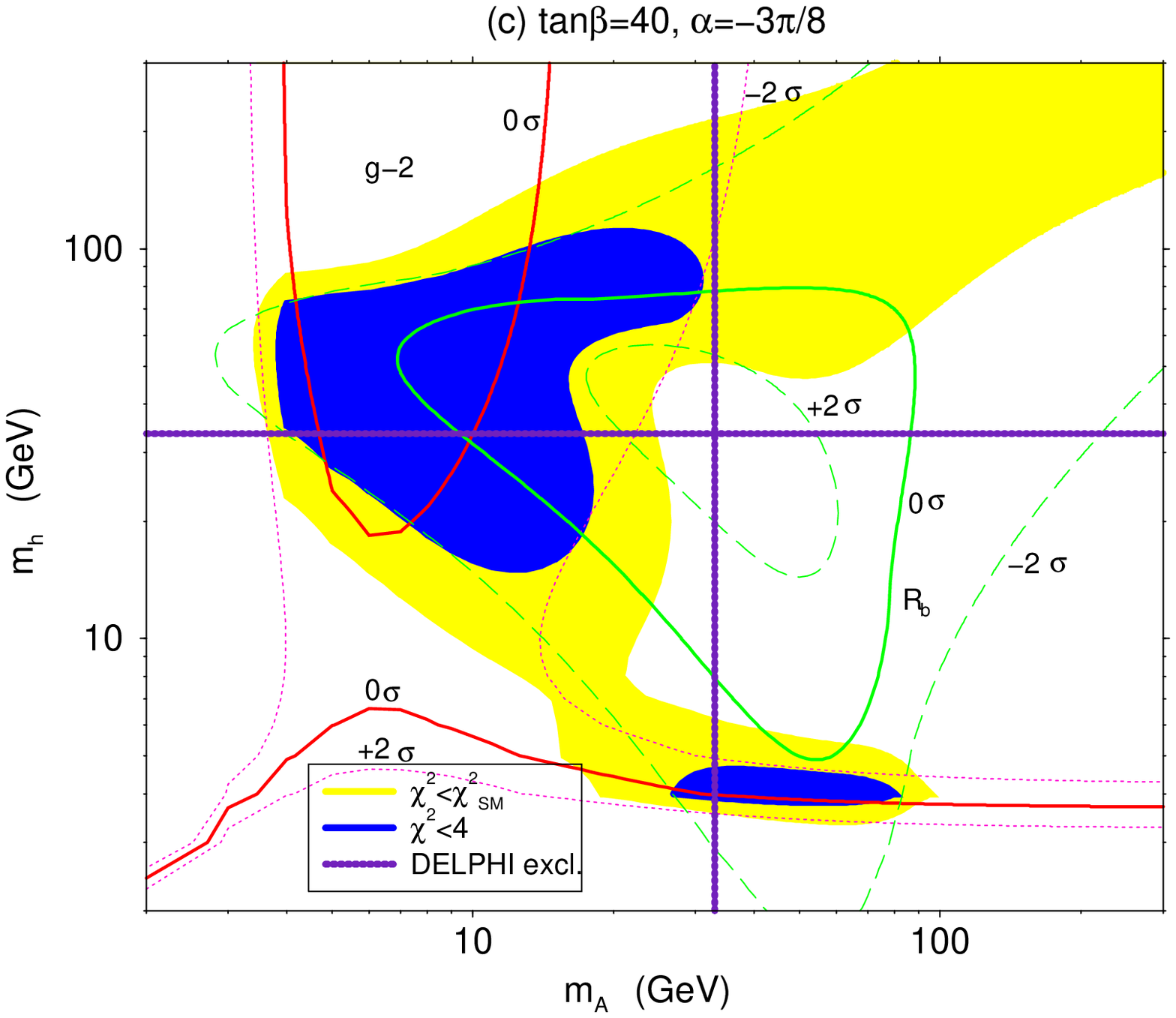}
\includegraphics[width=3.2in]{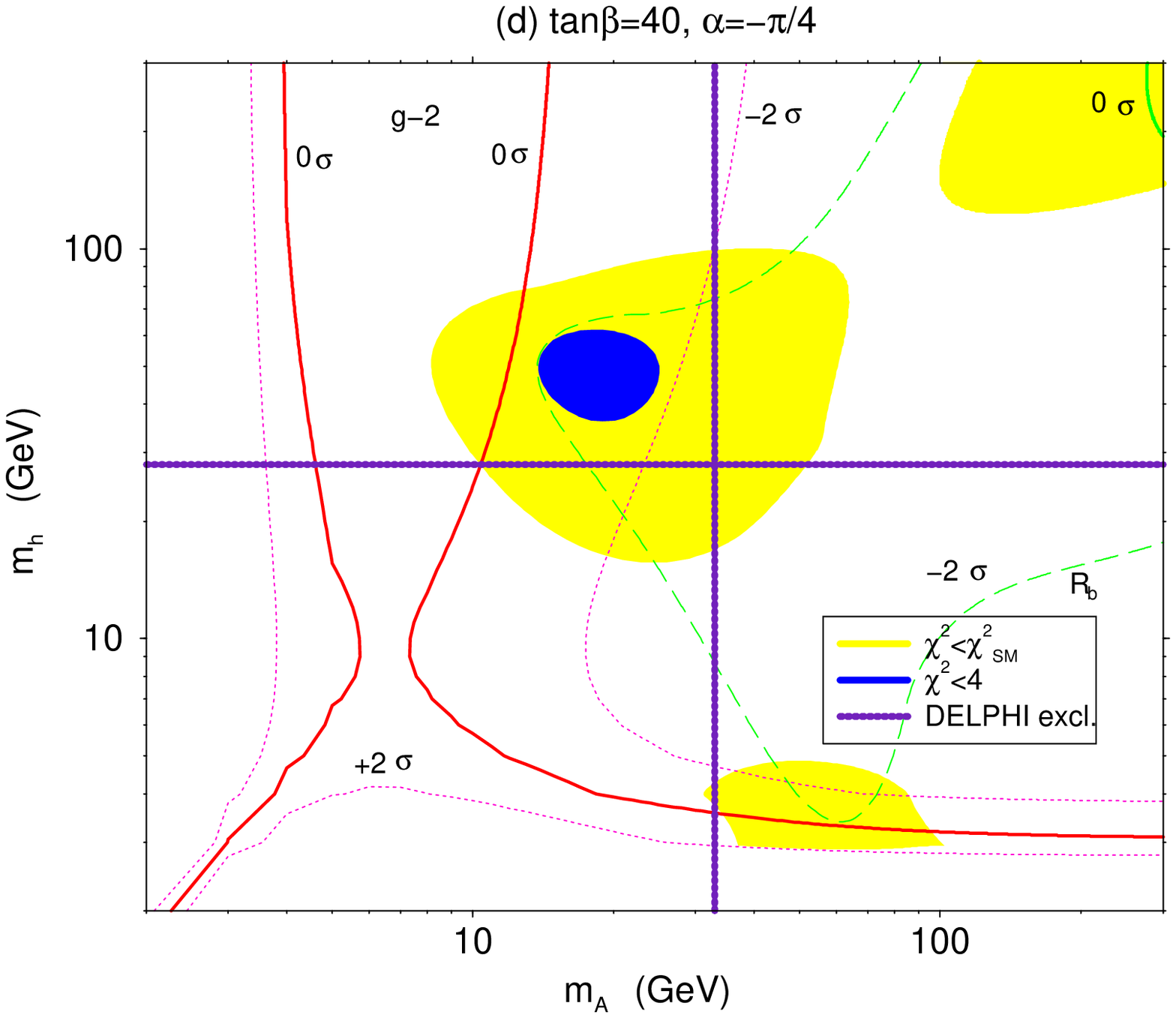}
\caption{\small
The $2\sigma$ allowed regions in the $(m_{\!\scriptscriptstyle A},\;m_h)$ plane due to the
constraints of $a_\mu$ and $R_b$ for $\tan\!\beta=40$.  
The smaller dark region is where
the total $\chi^2$ is less than 4 while the lighter region is where the
total $\chi^2$ is less than the $\chi^2({\rm SM})=10.3$.  Here the lower mass
limits on $m_{\!\scriptscriptstyle A}$ and $m_h$ from DELPHI are shown.
} 
\label{fig_4}
\end{figure}

For instance, the DELPHI exclusion we used in Fig.~\ref{fig_2} obviously kills 
quite a part of the above solutions. At $\tan\!\beta=58$, the
lower bound on $m_{\!\scriptscriptstyle A}$ is actually about
$40\,\mbox{GeV}$. It is particularly interesting to check the case of
a relatively small $\tan\!\beta$. We show in Fig.~\ref{fig_4} the case of 
$\tan\!\beta=40$, in which the DELPHI  almost kills all solutions. In the
figure, only a tiny window survives at $\sin\!\alpha={3\pi\over 8}$.
This is, however, at a $m_h$ (as well as $m_{\!\scriptscriptstyle A}$) 
value too small to survive the OPAL exclusion discussed below. From the 
same study by DELPHI, the Yukawa processes were also used to impose 
bounds on the $m_h$\cite{delphi}. Note that the plots in the 
Ref.~\cite{delphi} give bounds on Higgs masses versus the $b-\tau$ coupling 
enhancement factor, which is simply $\tan\!\beta \sqrt{B(A\to b\bar b,
\;\tau^+ \tau^-)}$
for the pseudoscalar case, but has an added $|\sin\!\alpha|$
dependence for the case of the scalar $h$. In the range of
$m_{\!\scriptscriptstyle A}$ values of interest, however, the
$m_h$ bound is typically superseded by the OPAL exclusion.
Note that the $a_\mu$ solution with
$m_h$ at a few $\mbox{GeV}$ is also inadmissible, from the 
consideration of $\Upsilon$-decay\cite{ours} and otherwise. 
Figure \ref{fig_4} also illustrates that a positive value of $\alpha$, around 
${3\pi\over 8}$ tends to give $a_\mu$-$R_b$ solutions with the
largest $m_{\!\scriptscriptstyle A}$.  This is mainly a result of
the $a_\mu$ constraint. At a fixed $m_h$, the contribution of the scalar
through the 2-loop Barr-Zee diagram to $a_\mu$ is suppressed by 
$|\sin\!\alpha|$, and thus allows a larger $m_{\!\scriptscriptstyle A}$. A
slight asymmetry in the cases of positive and negative  $\alpha$
values comes in as a result of the different way the 1-loop 
contributions go.

We should also point out that we find no solutions to the $a_\mu$-$R_b$ fit 
at all for much lower $\tan\!\beta$. The $a_\mu$ solutions shown in 
Fig.~\ref{fig_2}
simply produce a $m_h$-$m_{\!\scriptscriptstyle A}$ splitting 
too large to accommodate the $R_b$ constraint.  Therefore, the solutions to the
$a_\mu$-$R_b$ fits start to emerge as  $\tan\!\beta$ approaches a
large enough value, not much below 40. To get solutions with high
enough masses for the Higgs bosons so as to survive the exclusion limits
from the LEP searches, one will have to get to a higher and higher
$\tan\!\beta$ value.  To check the details, we first turn to the
powerful exclusions from OPAL.

\begin{figure}[th!]
\includegraphics[width=3.2in]{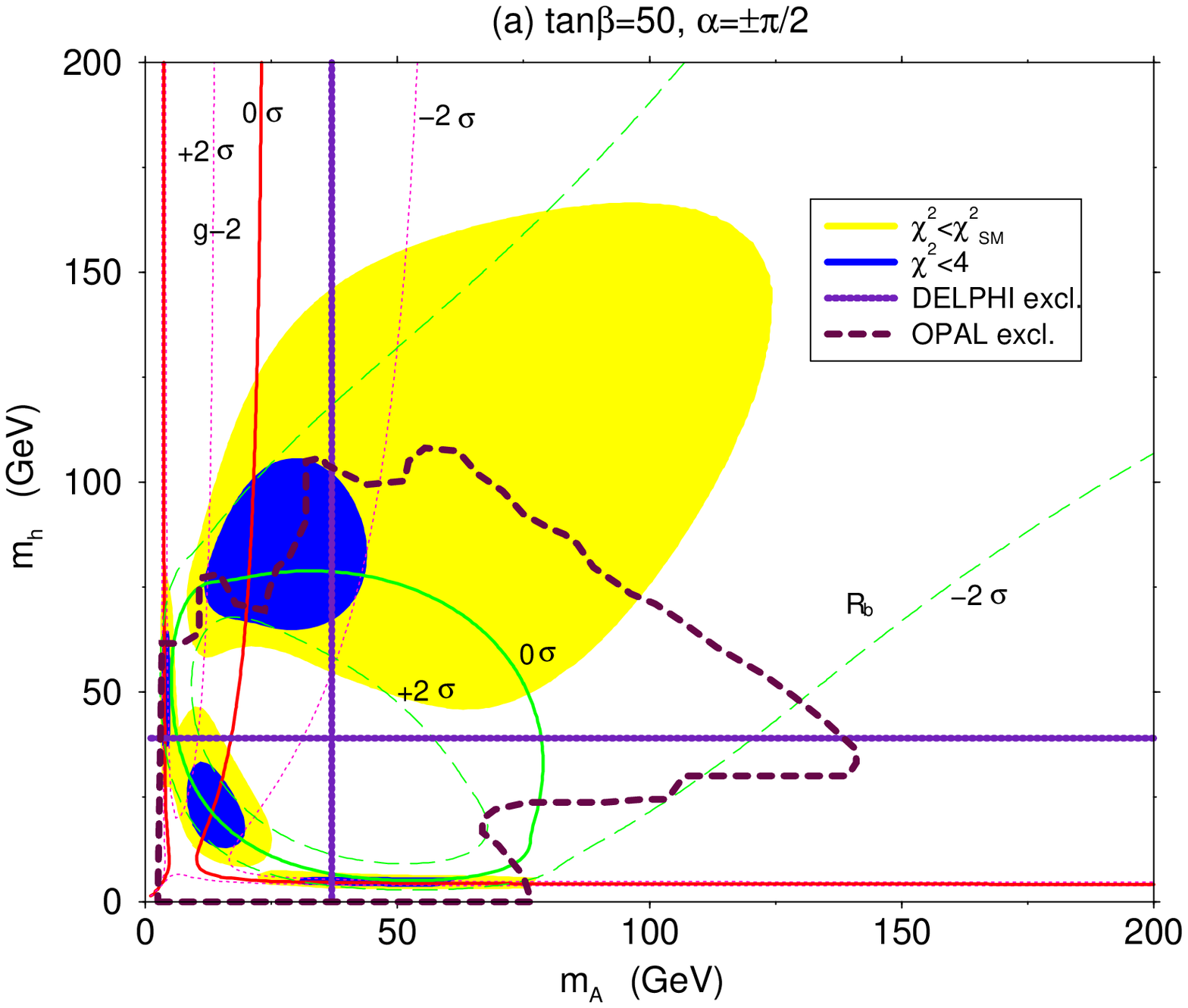}
\includegraphics[width=3.2in]{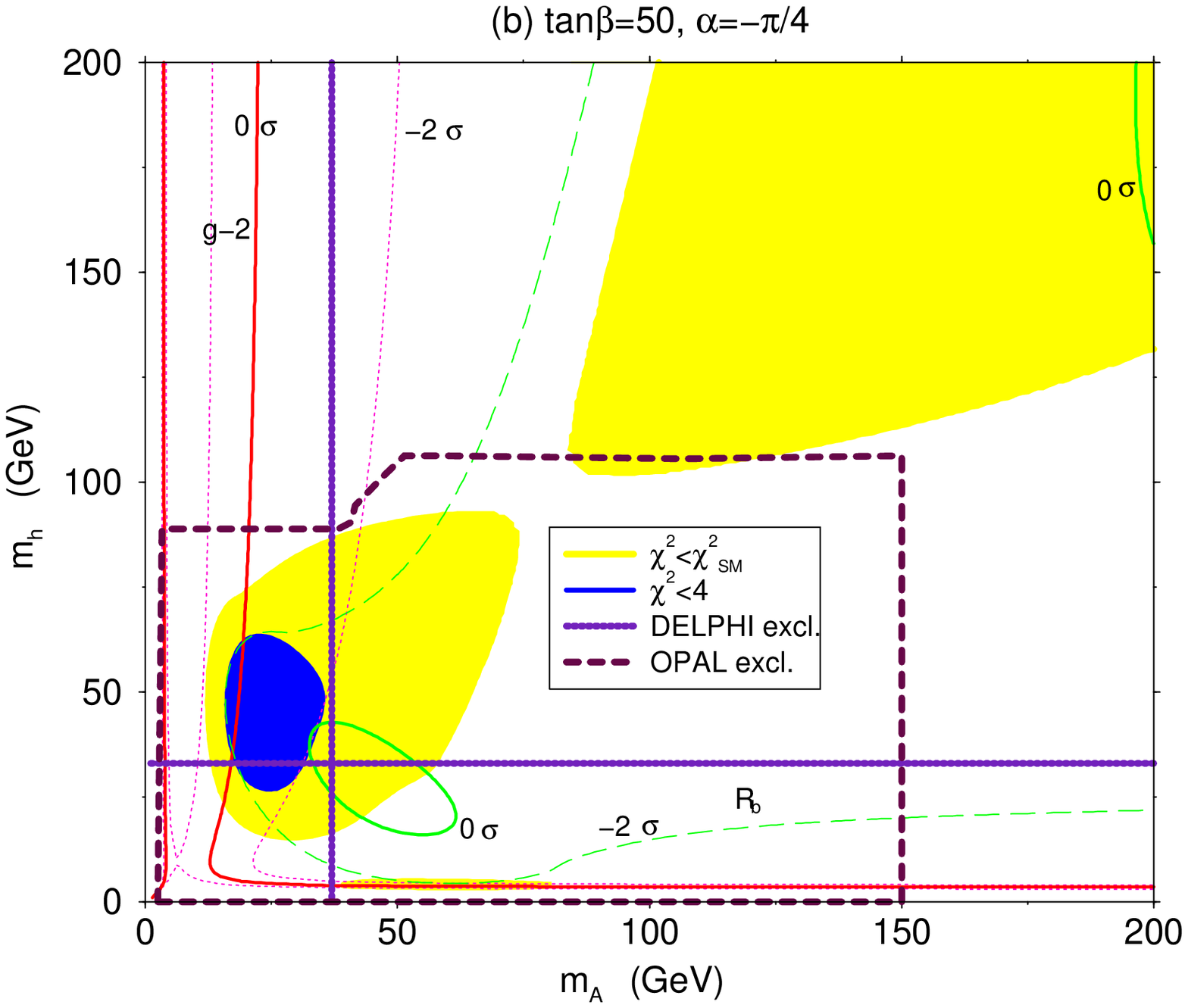}

\vspace{0.2in}

\includegraphics[width=3.2in]{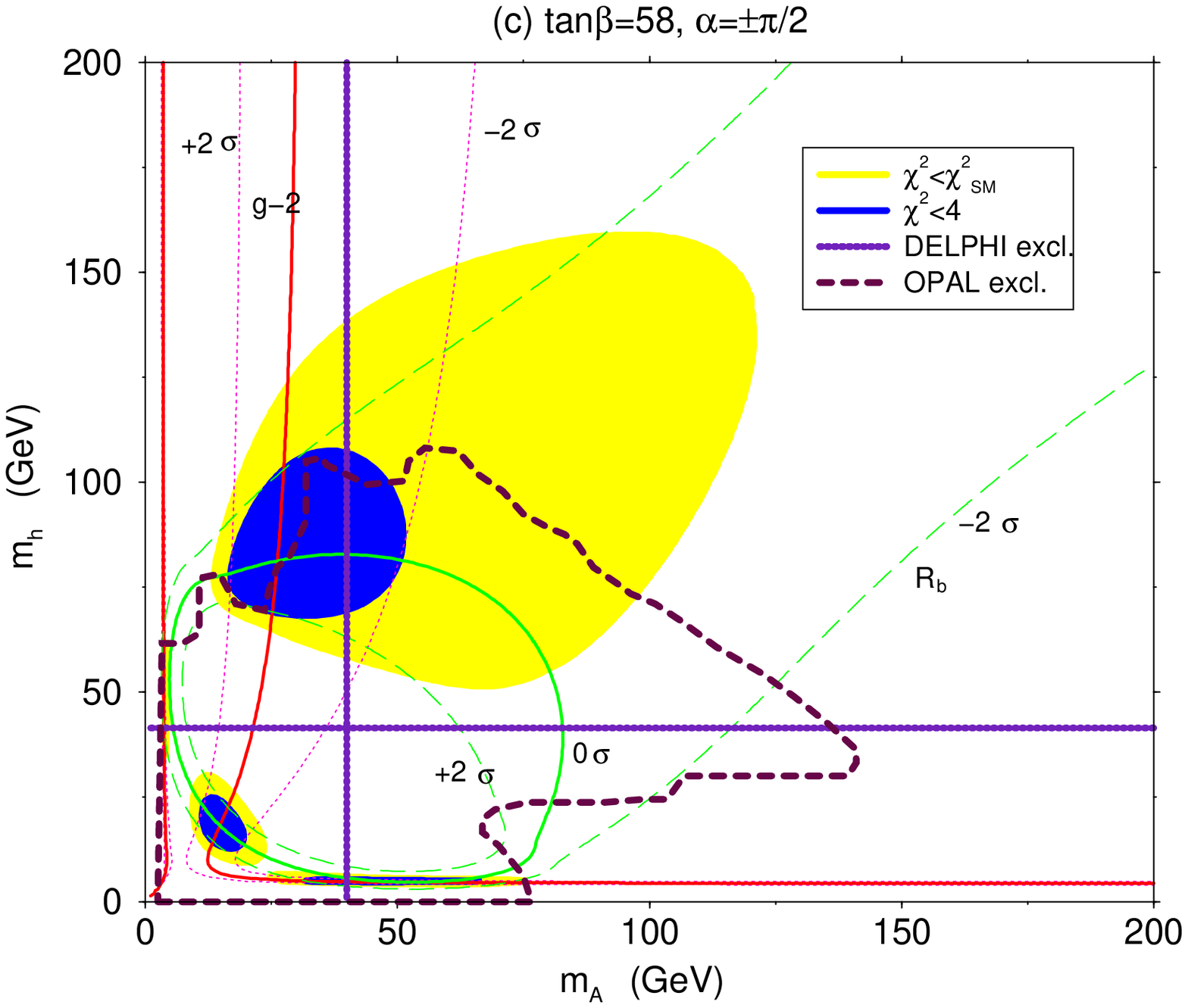}
\includegraphics[width=3.2in]{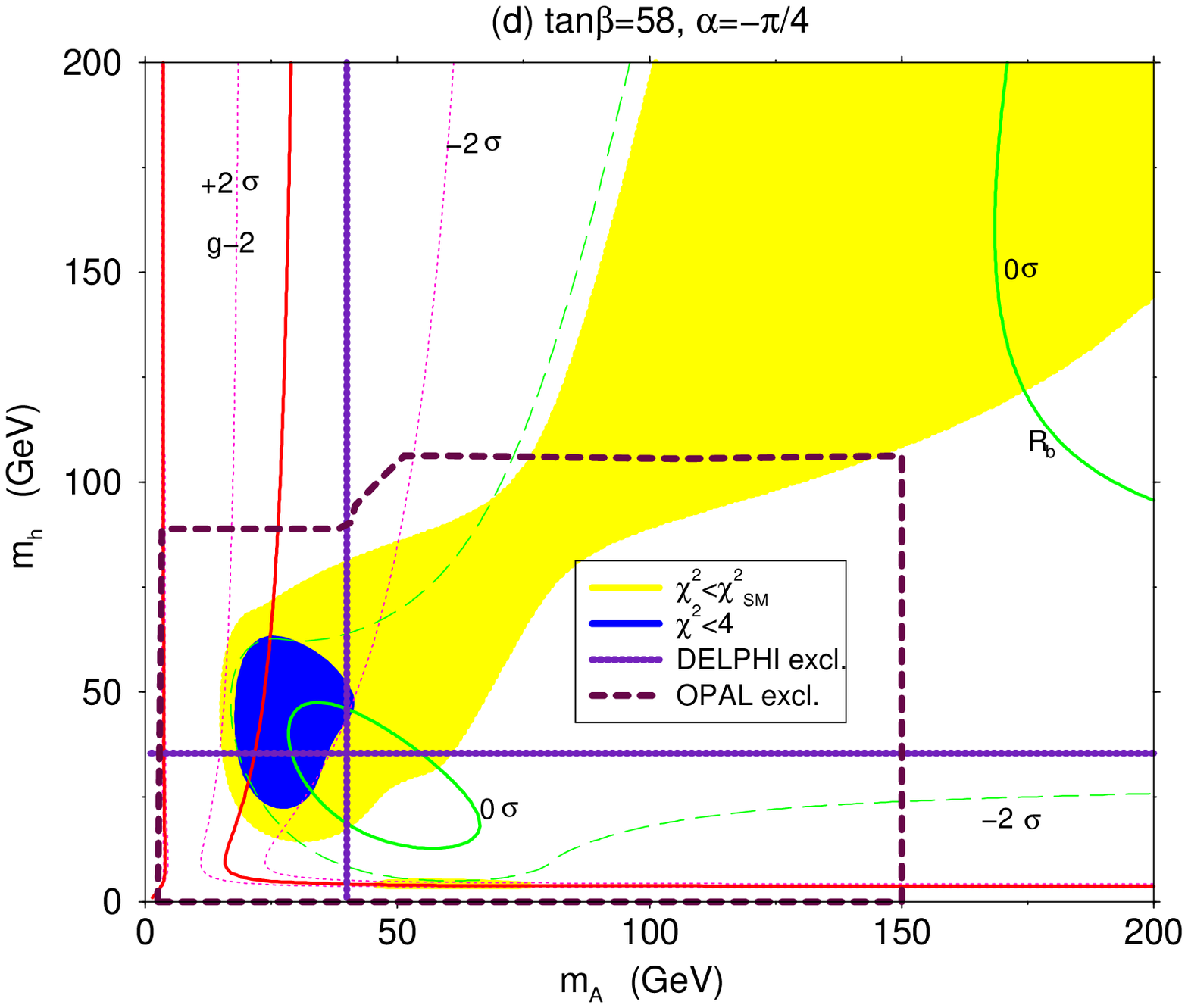}
\caption{\small
The $2\sigma$ allowed regions in the $(m_{\!\scriptscriptstyle A},\;m_h)$ plane due to the
constraints of $a_\mu$ and $R_b$ for $\tan\!\beta=50$ (a,b) and for 
$\tan\!\beta=58$ (c,d).
The smaller dark region is where
the total $\chi^2$ is less than 4 while the lighter region is where the
total $\chi^2$ is less than the $\chi^2({\rm SM})=10.3$.  Here the lower mass
limits on $m_{\!\scriptscriptstyle A}$ and $m_h$ from DELPHI, and the excluded region from 
OPAL are shown.
\label{fig_5}
} 
\end{figure}

Based also on the LEP data, OPAL has been publishing Higgs-search analyses 
specifically focused on the 2HDM. Here, we
use the most recent results available \cite{opal,pamela}. The 
results exclude a region at the lower left corner of the  
$m_h$-$m_{\!\scriptscriptstyle A}$ plane for each specific value 
of the mixing angle $\alpha$ (four explicitly shown). As presented in 
Ref.\cite{opal}, however, the results are not $\tan\!\beta$ specific. We 
show in Fig.~\ref{fig_5}
 the  solution regions from $a_\mu$-$R_b$ with the OPAL 
and the above mentioned DELPHI excluded regions superimposed. 
Here, we show the cases of $\tan\!\beta=50$ and $58$ for two $\alpha$ 
values, $-{\pi\over 4}$ and $\pm{\pi\over 2}$. The latter are chosen as
they are among the $\alpha$ values for which the OPAL paper gives the
explicit exclusion. The exclusion for non-specific $\alpha$ values 
only presents a substantially weakened result to be of interest here. The 
two $\alpha$ values are also close to or within the range of value 
for an optimal $a_\mu$-$R_b$ fit, as illustrated in Fig.~\ref{fig_3} above.
 
In fact, the OPAL exclusions are based on the $\tan\!\beta$ value
in the range $1-58$, given only at four values of $\alpha$, 
which we adopted here for the plots. An excluded point is one excluded 
at all value of $\tan\!\beta$ within the range. The excluded ranges may hence 
be extended at each specific value of $\tan\!\beta$, with more detailed
analysis of the data\cite{pamela2}. In fact, one would expect the enhanced
couplings at a larger $\tan\!\beta$ generally push the exclusion regions 
towards higher masses. One can see in the plots that the $a_\mu$-$R_b$ 
solution regions are largely cut off by the presented OPAL exclusions in 
general. Actually, nothing survives in the illustrated plots for 
$\tan\!\beta=50$ [{\it cf.} plots (a) and (b) of Fig.~\ref{fig_5}]; and it is also 
quite obvious that the same is true for the case of $\tan\!\beta=40$. Figures 
\ref{fig_4} and \ref{fig_5} together clearly illustrate the general trend of  how 
the increase in $\tan\!\beta$ gives better results. While it does not do much 
for the case of $\alpha=-{\pi\over 4}$, at $\alpha=\pm{\pi\over 2}$, the 
admissible $m_h$ and $m_A$ values are pushed to be high enough to escape 
the OPAL exclusion at  $\tan\!\beta>50$ [{\it cf.} plot (c) 
of Fig.~\ref{fig_5}].  
However, we are not bold enough to say for sure if no solution survives 
at $\tan\!\beta=50$  though. The optimal case giving the largest $m_h$ 
value for the $a_\mu$-$R_b$ is likely to be around $\alpha=-{3\pi\over 8}$,
at which the present result of OPAL did not show its greatest strength. 

We pick $\tan\!\beta=58$ for the above detailed illustration of the
 $a_\mu$-$R_b$ fits because it gives the most favorable case within the limit of 
a direct application of the OPAL exclusion. There is apparently a surviving
window in the parameter space.  Combining our present study with a refined 
version of the OPAL analysis to focus on our $a_\mu$-$R_b$ solution regions, 
especially the upper blots as shown in plots (a) and (c) of Fig.~\ref{fig_5}, 
will certainly be very interesting. Exclusions have to be checked for each
specific value of $\tan\!\beta$ and that of $\alpha$. Such a study will further 
narrow down the survival parameter space regions of the 2HDM, especially 
with further improved exclusion limits. We are told that the LEP data actually 
allows such an improvement\cite{pamela2}. A very tantalizing question is if 
the current constraints are actually strong enough to kill the model altogether!  
 
The OPAL analysis is limited to $\tan\!\beta$ value at or below $58$, 
while the DELPHI analysis stops at $100$. The very large 
$\tan\!\beta$ region is theoretically unfavorable and may provide
practical problems due to the much enhanced $b$-quark Yukawa
coupling, which signals a breakdown of the perturbative treatment.
Nevertheless, we will include some results from such uncomfortably
large $\tan\!\beta$ values, and urge the OPAL group to push on a bit
further in their analysis.

\begin{figure}[th!]
\includegraphics[width=3.2in]{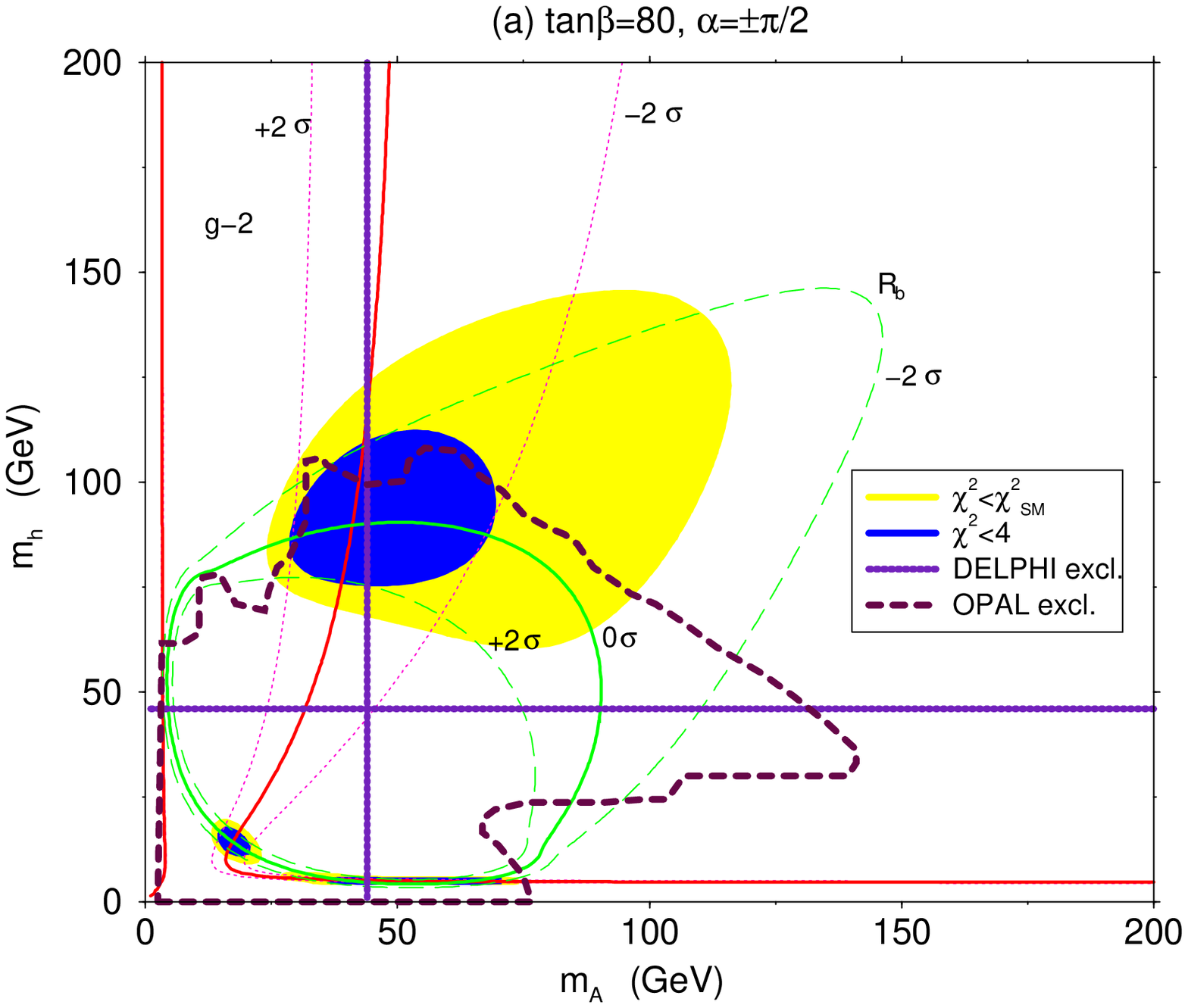}
\includegraphics[width=3.2in]{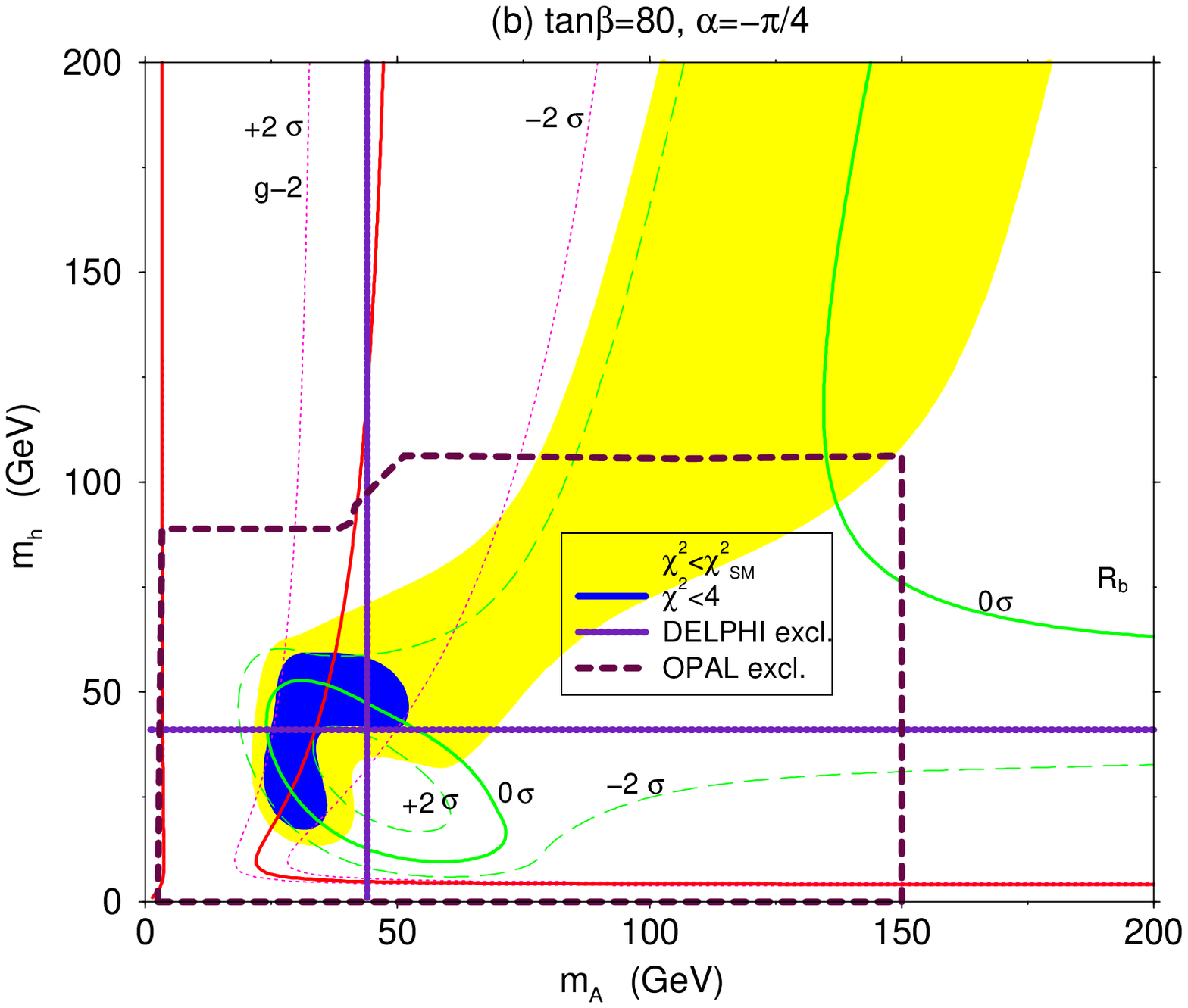}

\vspace{0.2in}

\includegraphics[width=3.2in]{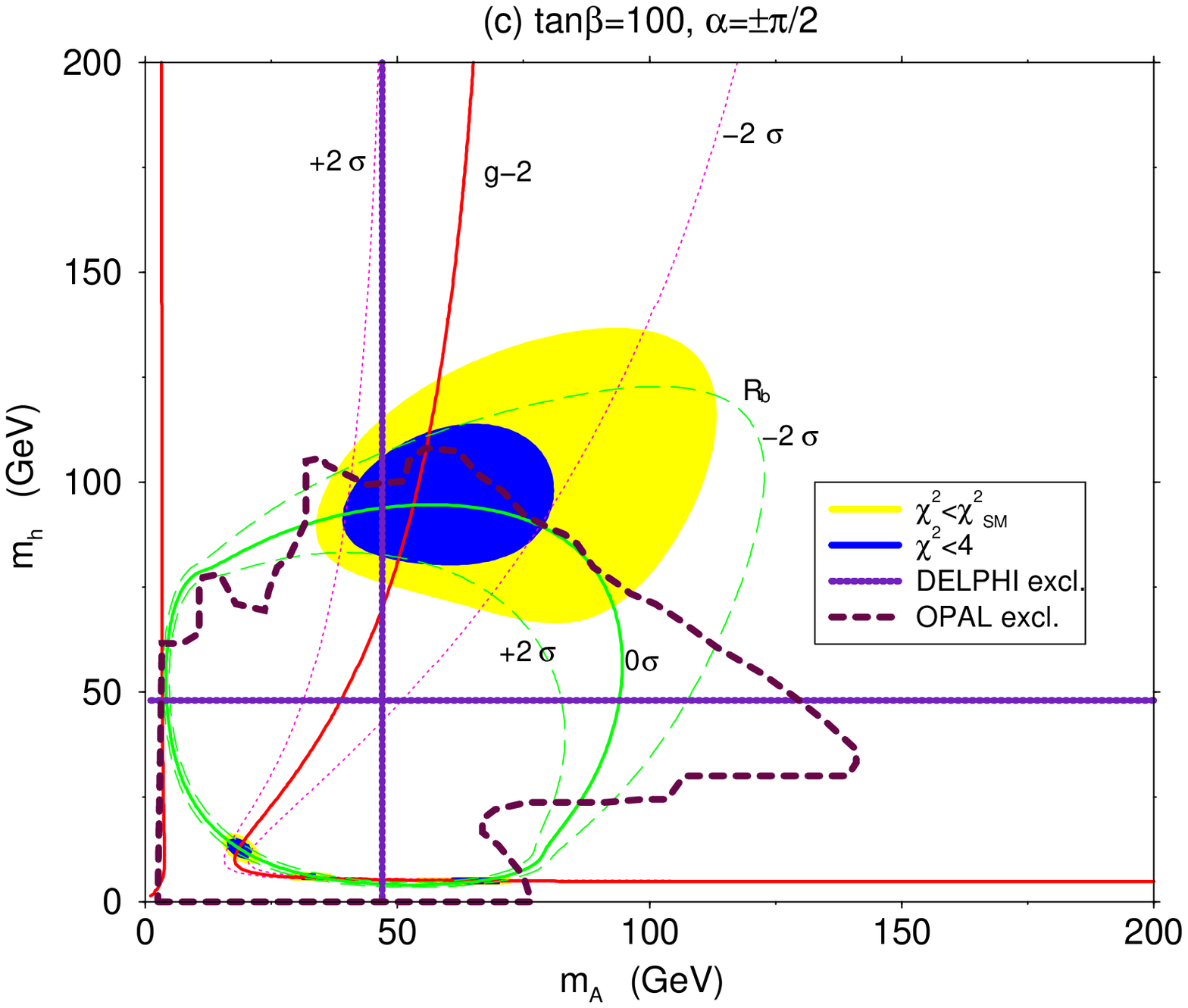}
\includegraphics[width=3.2in]{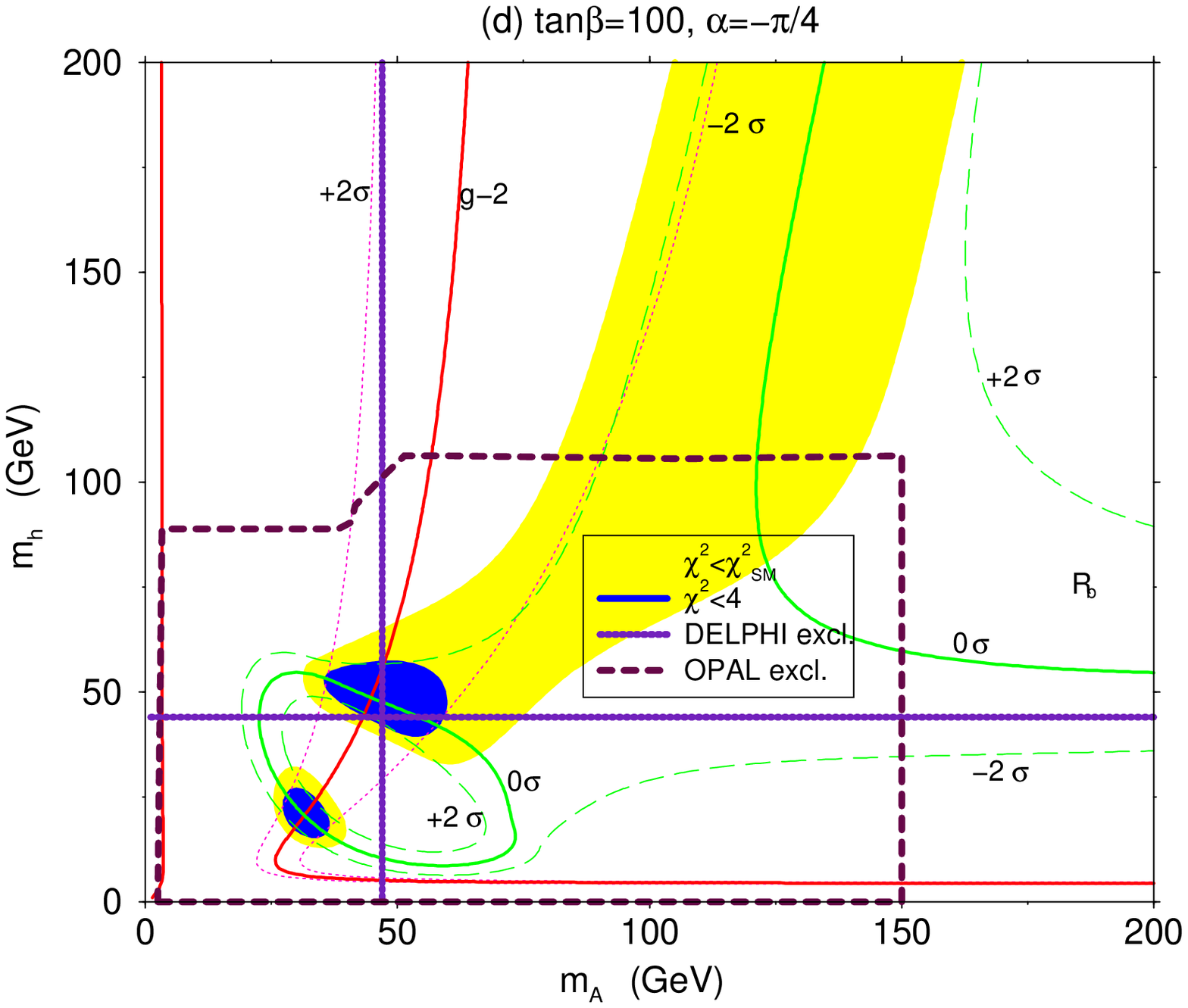}
\caption{\small
Same as Fig.~\ref{fig_5}, but for $\tan\!\beta=80$ (a,b) and for
$\tan\!\beta=100$ (c,d). Note that the OPAL exclusion regions put in is
actually for $\tan\!\beta\leq 58$, hence no longer directly applicable here. In lack
of the corresponding applicable results, they are kept here for reference. 
\label{fig_6}
} 
\end{figure}

\begin{figure}[th!]
\includegraphics[width=6in]{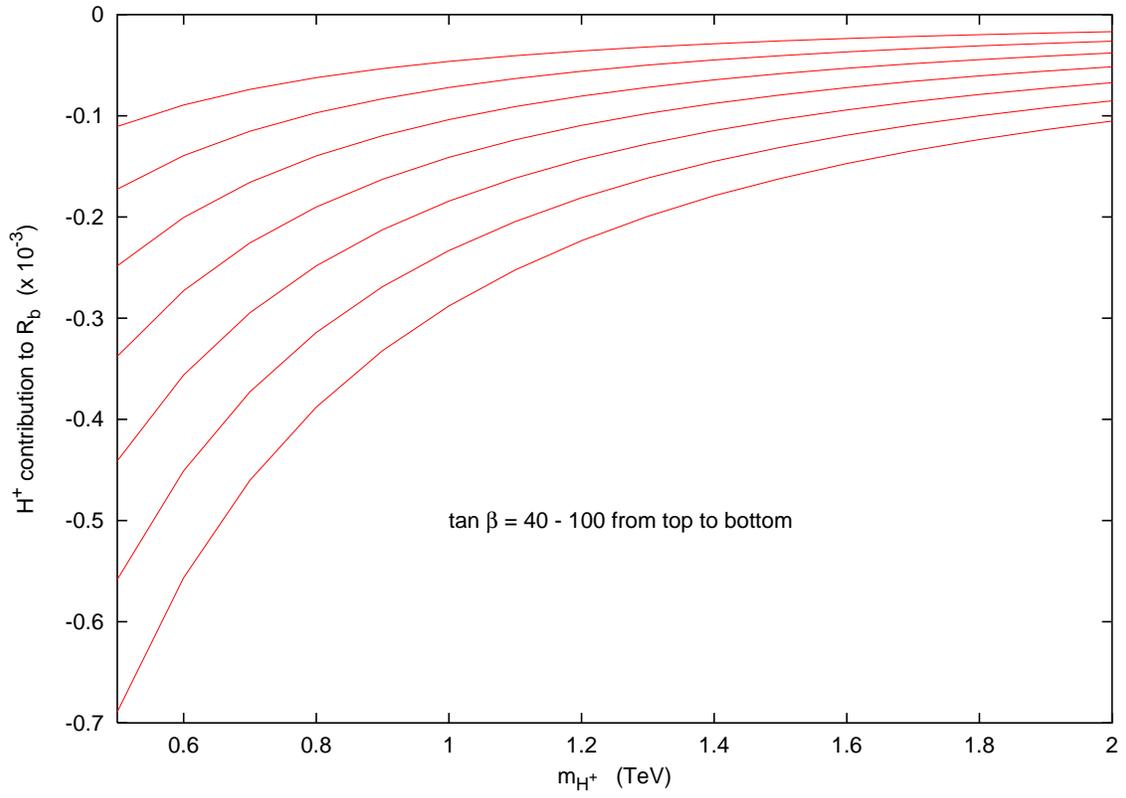}
\caption{\small
The charged-Higgs contribution to $R_b$ in units of $10^{-3}$ for 
$\tan\!\beta=40-100$ (from the top to the bottom).
\label{fig_7}
} 
\end{figure}

For $\tan\!\beta>58$, we again illustrate some results for 
$\alpha=-{\pi\over 4}$ and $\pm{\pi\over 2}$ in Fig.~\ref{fig_6}, 
in which we still
put in the (no longer exactly valid) OPAL exclusion from the 
$\tan\!\beta=1-58$ scan for reference. For the $\alpha={-\pi\over 4}$ 
case, the $a_\mu$-$R_b$ solution regions never rise above 
$m_h=60\,\mbox{GeV}$ (as also for the smaller  $\tan\!\beta$ cases),
and partially excluded by DELPHI. There seem to be good
enough reasons to believe that such low Higgs mass regions
should be excluded by the available LEP data if an analysis along
the OPAL line is performed.  The $\alpha=\pm{\pi\over 2}$ case is 
better. The surviving region seen at $\tan\!\beta=58$ moved 
further to the right, towards larger $m_{\!\scriptscriptstyle A}$,
and further up a bit for larger $m_h$.  The rise in $m_h$ actually
more or less saturates at $\tan\!\beta=100$, and falls for even
larger values of $\tan\!\beta$.
Such a region still lives at the boundary of the DELPHI exclusion,
but would still have a surviving part even if the OPAL exclusion
can still be imposed. The region is, however, shrinking with
increases in $\tan\!\beta$, due to a more fine-tuned $a_\mu$-$R_b$ 
solution. One should also bear in mind that the (upward) shifting,
and hence slight enlargement, of this solution region with a further
increase in the $m_{\!\scriptscriptstyle H^\pm}$ value. For this
purpose, we give the charged-Higgs contribution to $R_b$ in Fig.~\ref{fig_7}.

Let us summarize our results so far.
We have seen that combining the suggested requirement of producing
a definite positive contribution to $a_\mu$ while keeping a limited
deviation from the SM $R_b$ result is an extremely stringent
constraint on the parameter space of the 2HDM. When the 
available direct experimental search results from the LEP experiments
are further implemented, there is at most a tiny window of parameter
space that can survive. The apparently surviving region from the 
above discussions is restricted to very large values of $\tan\!\beta$.
In fact, it may be already uncomfortably large, inviting the problem
of the perturbativity of the Yukawa coupling of the $b$ quark. 
As for the mixing angle $\alpha$, it is being pushed close to the
$-{\pi\over 2}<\alpha<-{3\pi\over 8}$ region.  All these are based on 
a strong mass splitting between the charged Higgs and the light scalars, 
with the pseudoscalar $A$ lying typically below $80\,\mbox{GeV}$ and the 
scalar  $h$ below $140\,\mbox{GeV}$.

Recall that we have essentially decoupled the heavy Higgs boson
$H$ in the above analysis. We promised to justify this as a 
physics requirement. We first noted that a not too heavy $H$ would
more or less add to the effect of the other real scalar $h$ in the
contributions to $a_\mu$ and $R_b$. So, we expect it to ask for a
larger mass splitting when fitting $a_\mu$ is concerned, but a
smaller mass splitting to fit $R_b$. In another word, further
tightening of the apparent solution window. In the section below,
we will show that fitting another precision EW parameter, the $\rho$ parameter,
actually does require sending $m_{\!\scriptscriptstyle H}$ to a very large
value, indeed well beyond $m_{\!\scriptscriptstyle H^\pm}$.

\section{The $\rho$ parameter constraint}
The parameter $\rho$ was introduced to measure the relation between the 
masses of $W^\pm$ and $Z^0$ bosons.  In the SM 
$\rho \equiv M_{\! \scriptscriptstyle W}^2/M_{\! \scriptscriptstyle Z}^2 
\cos\!^2\theta_{\! \scriptscriptstyle W}=1$ at tree-level.
However, the $\rho$ parameter receives contributions from the SM corrections 
and from new physics.  The deviation from the SM prediction is usually described 
by the parameter $\rho_{\scriptscriptstyle 0}$ defined by\cite{langacker}
\begin{equation}
\rho_{\scriptscriptstyle 0} \equiv
 \frac{M_{\! \scriptscriptstyle W}^2}{\rho \, M_{\! \scriptscriptstyle Z}^2 
\cos^2 \!\theta_{\! \scriptscriptstyle W}} \;,
\end{equation}
where the $\rho$ in the denominator absorbs all the SM corrections, 
including the corrections from the top quark and the SM Higgs boson.
By definition, $\rho_{\scriptscriptstyle 0}=1$  in the SM. Sources of new 
physics that contribute to $\rho_{\scriptscriptstyle 0}$ can be written as
\begin{equation}
\rho_{\scriptscriptstyle 0} = 1 + \Delta \rho_{\scriptscriptstyle 0}^{\rm new} \;,
\end{equation}
where $\Delta \rho_{\scriptscriptstyle 0}^{\rm new}= \Delta \rho^{\mbox{\tiny 2HDM}} -          \Delta \rho^{\mbox{\tiny SM-Higgs}}$ in our case.
Note that since the two-doublet Higgs sector (in the 2HDM) is employed 
here to replace the SM Higgs, the latter contribution to $\Delta \rho$ has to 
be subtracted out.

The most recent reported value of $\rho_{\scriptscriptstyle 0}$ is \cite{pdg2002}
\begin{equation}
\label{rho0}
\rho_{\scriptscriptstyle 0} = 1.0004 \pm 0.0006,\quad 
\mbox{(with $M_{\! \scriptscriptstyle H_{\rm SM}}$ fixed at 115 GeV)} \;.
\end{equation}
In terms of new physics the constraint becomes:
\begin{equation}
\Delta \rho_{\scriptscriptstyle 0}^{\rm new} = 0.0004 \pm 0.0006 \;.
\end{equation}
In 2HDM $\Delta \rho$ receives contributions from all Higgs bosons given by
\cite{hunter,maria-ew}
\begin{eqnarray}
\label{rho}
\Delta \rho^{\mbox{\tiny 2HDM}} &=& 
\frac{\alpha_{\rm em}}{4\pi \sin^2\! \theta_{\! \scriptscriptstyle W}
 M_{\! \scriptscriptstyle W}^2} \, 
\Big[ \, F( m_{\! \scriptscriptstyle A}, m_{\! \scriptscriptstyle {H^+}} ) 
 + \cos^2\!(\beta -\alpha) \, [ F(m_{\! \scriptscriptstyle {H^+}},m_h) 
- F(m_{\! \scriptscriptstyle A}, m_h) ] 
 \nonumber \\ && 
 + \sin^2\!(\beta -\alpha) \, [ F(m_{\! \scriptscriptstyle {H^+}}, m_{\! \scriptscriptstyle {H}} ) 
 - F(m_{\! \scriptscriptstyle A},m_{\! \scriptscriptstyle {H}})] \,\Big ]
 \nonumber \\
&& + \cos^2\!(\beta-\alpha) \Delta \rho^{\mbox{\tiny SM}}\!(m_{\! \scriptscriptstyle {H}}) 
   + \sin^2\!(\beta-\alpha) \Delta \rho^{\mbox{\tiny SM}}\!(m_h) \;,
\end{eqnarray}
where
\begin{eqnarray}
F(x,y) &=& {1\over 8}x^2 +{1\over 8}y^2-{1\over 4}\frac{x^2 y^2}{x^2-y^2} 
\log\!\!\left({x^2\over y^2}\right) \; = F(y,x) \;,
\nonumber \\
\Delta \rho^{\mbox{\tiny SM}}\!(M) &=& -
\frac{\alpha_{\rm em}}{4\pi \sin^2\! \theta_{\! \scriptscriptstyle W}
 M_{\! \scriptscriptstyle W}^2} \, \left[   3 F(M, M_{\! \scriptscriptstyle W})
- 3 F(M, M_{\! \scriptscriptstyle Z} )  +{1\over 2} (M_{\! \scriptscriptstyle Z}^2
-M_{\! \scriptscriptstyle W}^2) \right ] \;.
\end{eqnarray}

Let us take a closer look into the implication of the formulas above.
First of all, we note that $\Delta \rho^{\mbox{\tiny SM}}\!(M)$ has a negative
value with magnitude increasing with $M$. As to be expected from above, the value 
is about $-0.0004$ at $M=115\,\mbox{GeV}$. It has a relatively mild variation, 
and does not go beyond $-0.005$ even as $M$ gets to $10\,\mbox{TeV}$.
In direct contrast, the other contributions to $\Delta \rho^{\mbox{\tiny 2HDM}}$
in the above formula are very sensitive to the masses involved. The $F(x,y)$ 
function is always positive, vanishes only at $x=y$, and increases with a 
faster
and faster rate with the splitting between $x$ and $y$. In a typical scenario
that is of interest here, we expect the pseudoscalar to be the lightest Higgs state
with a quite heavy charged Higgs. That makes the contribution from the
first term [involving $F( m_{\! \scriptscriptstyle A}, m_{\! \scriptscriptstyle {H^+}})$]
large; indeed of order $0.01$ for $m_{\! \scriptscriptstyle {H^+}}$ satisfying
the lower bound from $b\to s\,\gamma$ and $B-\overline{B}$ mixing. To get the 
required almost zero value of  $\Delta \rho^{\mbox{\tiny 2HDM}}$, we need some
negative contributions from the terms involving
$[ F(m_{\! \scriptscriptstyle {H^+}},m_h) - 
F(m_{\! \scriptscriptstyle A}, m_h) ]$
and  $[ F(m_{\! \scriptscriptstyle {H^+}}, m_{\! \scriptscriptstyle {H}} ) 
 - F(m_{\! \scriptscriptstyle A},m_{\! \scriptscriptstyle {H}})]$. And the 
solution
is obviously a fine-tuned one. Consider a case of degenerated Higgs real 
scalars,
$m_{\! \scriptscriptstyle {H}}=m_h=M$. The $(\beta-\alpha)$ dependence in
the formula is removed as the sine-square and cosine-square parts are
combined. We need 
$[ F(m_{\! \scriptscriptstyle {H^+}},M) - F(m_{\! \scriptscriptstyle A}, M) ]$
to have a negative value close in magnitude to that of
$F( m_{\! \scriptscriptstyle A}, m_{\! \scriptscriptstyle {H^+}})$. The former is 
obviously positive roughly when $M$ is closer to 
$m_{\! \scriptscriptstyle A}$ than $m_{\! \scriptscriptstyle {H^+}}$, and
negative when it is the other way round; and it can be larger than
$F( m_{\! \scriptscriptstyle A}, m_{\! \scriptscriptstyle {H^+}})$
only for $M>m_{\! \scriptscriptstyle {H^+}}$. The $(\beta-\alpha)$ dependence
comes back in the generic situation, with a mass splitting between the two
real scalars. Obviously, at least one of them, by definition $H$, has to be
heavy. However, 
they cannot be both heavy, because the $R_b$ constraint does not
allow only a light pseudoscalar giving a substantial contribution. Hence,
this additional requirement of a limited splitting
between $m_{\! \scriptscriptstyle A}$ and $m_h$ clearly suggests a 
large $m_{\! \scriptscriptstyle {H}}$, typically 
larger than $m_{\! \scriptscriptstyle {H^+}}$. The larger the
$m_{\! \scriptscriptstyle {H}}$ value, the smaller the 
$\sin^2\!(\beta-\alpha)$ is required. Admissible solutions, though fine-tuned, 
can be obtained 
 so long as the required $\sin^2\!(\beta-\alpha)$ falls into the legitimate
interval. A large splitting between $m_{\! \scriptscriptstyle {H}}$ and $m_h$
also resulted in a very narrow range of admissible $(\beta-\alpha)$ values,
and hence the $\alpha$ values at a fixed $\tan\!\beta$ of interest.

\begin{figure}[th!]
\includegraphics[width=6in]{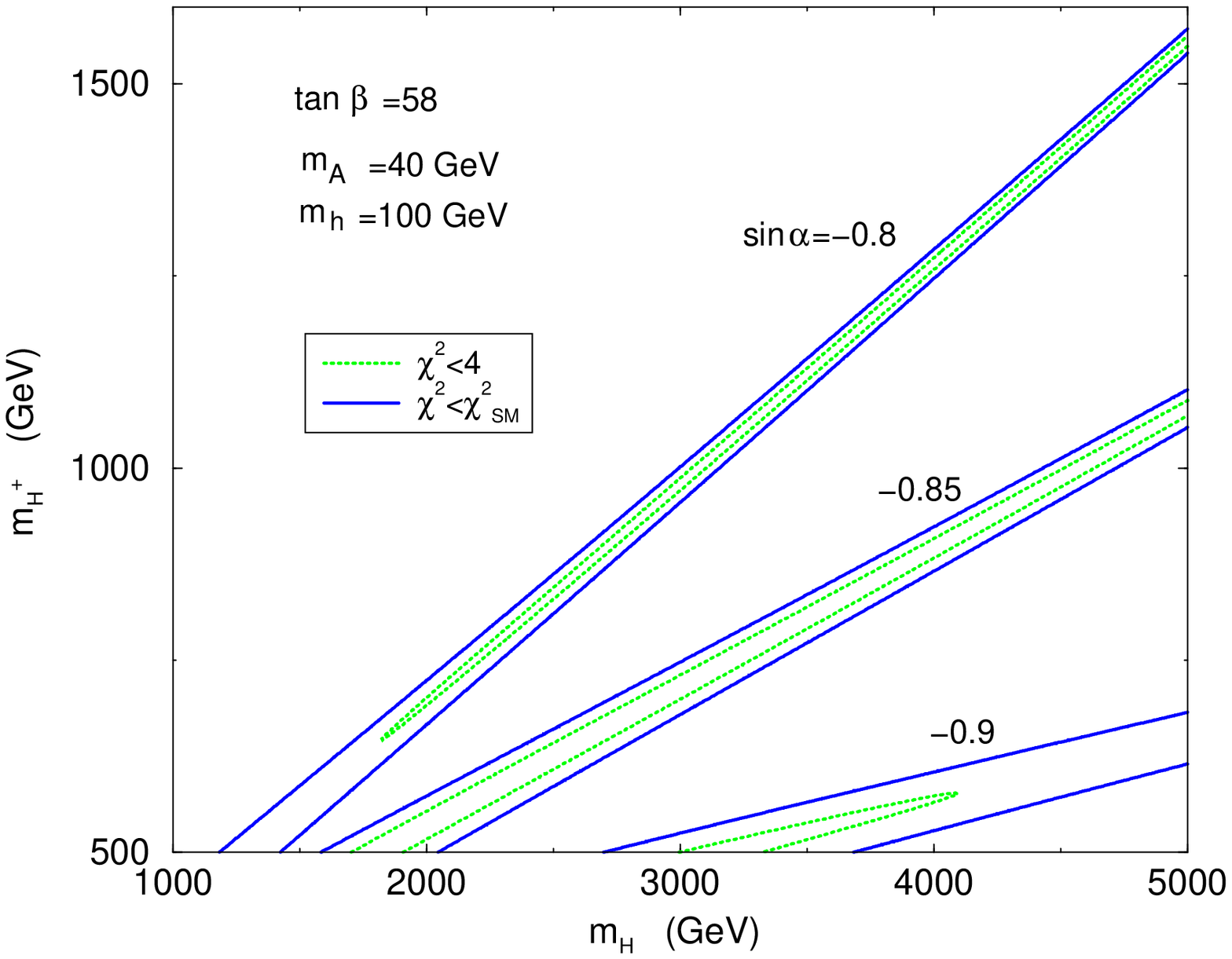}
\caption{\small
The allowed region in the plane of $(m_{\!\scriptscriptstyle H}, 
m_{\!\scriptscriptstyle H^+})$ 
due to constraints of $a_\mu$, $R_b$, and the $\rho$ parameter.  
Here the $m_{\!\scriptscriptstyle A}=40$ GeV and
$m_h=100$ GeV for $\tan\!\beta=58$ are chosen from the allowed 
darker  region in Fig.~\ref{fig_5}(c).
\label{fig_8}
} 
\end{figure}

Our numerical results corroborate well with the above analytical discussions.
We illustrative our discussion with a plot in Fig.~\ref{fig_8}. 
Here, we take a 
``surviving" solution point to the $a_\mu$-$R_b$ fits and perform
a further fitting together with the $\rho$ parameter by varying
$m_{\! \scriptscriptstyle {H}}$ and $m_{\! \scriptscriptstyle {H^+}}$. The
extremely fine-tuned nature of the solution is well-illustrated by the very
narrow $\chi^2$ bands. Moreover, $m_{\! \scriptscriptstyle {H}}$ 
is always more than twice of $m_{\! \scriptscriptstyle {H^+}}$ for an
$|\sin\alpha|$ larger than $0.8$. Note that $\sin\!\alpha$ of
$-0.8$ and $-0.92$ roughly correspond to an $\alpha$ value of
${-3\pi\over 10}$ and ${-3\pi\over 8}$, respectively. The basic features
remain if other solution points are taken, hence, we refrain
from showing more plots.

\section{Conclusions}
The requirement for the 2HDM II to give rise to the suggested
$+3\sigma$ deviation in the muon anomalous magnetic moment
demands a light pseudoscalar, preferably around $40\,\mbox{GeV}$
for $\tan\!\beta \le 58$. The FCNC constraints, in particular the 
$b\to s\,\gamma$, 
require a heavy charged Higgs beyond $400\,\mbox{GeV}$. This
large mass splitting in the Higgs mass spectrum is difficult to be accommodated
by the precision EW measurements. In particular, the $\rho$ parameter
constraint then admits only very fine-tuned solutions, with cancellation
from opposite contributions good to one in a hundred, favoring heavy
real scalars. The fine-tuned nature of the solutions, while making
many physicists uncomfortable with the model, is not in itself a good
enough reason to pronounce the death of the model. If one has
reasons to be confident about the correctness of the model, one
would say that the available constraints are just strong enough to pin
down for us the values of the unknown model parameter. A good 
example of such a situation is given by the pinning down of the top
mass value from precision EW data prior to the experimental discovery
of the top quark.

Nevertheless, we have not had much of a reason to believe in the 
correctness of the 2HDM. In fact, in our analysis here, we focus
more on the simultaneous fits to $\Delta a_\mu$ and $R_b$. Having
only the pseudoscalar contribution dominated the corrections to $R_b$
is fatal. Hence, we require a relatively light $m_h$, while pushing
$m_{\! \scriptscriptstyle {H}}$ to way beyond 
$m_{\! \scriptscriptstyle {H^+}}$ in order to satisfy the $\rho$
parameter. Even then, not much of a solution to the
$a_\mu$-$R_b$ fits survives the direct search exclusion.
Here, we require a total $\chi^2$ of 4 or smaller for the 
$a_\mu$-$R_b$ fits to claim a good solution.
For roughly $\tan\!\beta<40$, no solution to $a_\mu$-$R_b$ fits 
survives the DELPHI exclusion. In fact, there is no solution to the
$a_\mu$-$R_b$ fit for a $\tan\!\beta$ value quite a bit smaller
than 40.  For $\tan\!\beta$ value from around 40 to a bit
beyond 50, solutions surviving the DELPHI exclusions exist, but
only to be killed by the OPAL exclusions. For even larger
$\tan\!\beta$, the $a_\mu$-$R_b$ solutions surviving both the
DELPHI and OPAL exclusions started to emerge. This is
very much restricted to an $\alpha$ value in the range 
$-{\pi\over 2}<\alpha<-{3\pi\over 8}$. Solution regions shrink fast 
outside the range as the $m_h$ values given by the $a_\mu$-$R_b$ 
fits drop towards OPAL exclusion bound.  

In summary, under the strong restriction of the available constraints,
we show that only a very tiny window of apparent solutions exist 
close to the limit of $\tan\!\beta \le 58$. If the OPAL group could tailor
their analysis of the LEP data to focus on the apparent solution window
as shown here, we would be able to have a more definite conclusion. 
It looks to us that the solution window will be shut down quite substantially.
So, we have ``almost enough" constraints to kill the model altogether. For
$\tan\!\beta$ beyond the $58$ limit, our hands are tied at the moment
by the unavailability of the strong LEP exclusion results as presented
by OPAL. The very large  $\tan\!\beta$ values certainly make many
of us uncomfortable though. It is theoretical undesirable as limited by 
the blowing up of the bottom Yukawa couplings. 

\section*{\bf Acknowledgments}
We thank D.~Chang, C.~Kao, W.-Y.~Keung, and A.~I.~Sanda for valuable 
discussions. We are also in debt to P.~Ferrari from OPAL and M.~Boonekamp 
from DELPHI for clarification about their results.
This research was supported in part by the National Science Council
grant number NSC91-2112-M-008-042 (O.K.) and by the NCTS.

Otto Kong thanks the hospitality of National Center for Theoretical Sciences, 
Taiwan during the early phase of the study, and that of KEK Theory Group, 
Japan where the final draft is finished.

\newpage

\begin{thebibliography}{99}
\bibitem{BNL}
H.N. Brown {\it et al.}, Phys. Rev. Lett. {\bf 86}, 2227 (2001);
G.W. Bennett {\it et al.}, Phys. Rev. Lett. {\bf 89}, 101804 (2002),
Erratum-ibid. {\bf 89}, 129903 (2002).

\bibitem{amudata}
M. Davier, S. Eidelman, A. Hocker, and Z. Zhang, 
Eur. Phys. J. {\bf C27}, 497 (2003);
E. de Rafael, e-Print Archive: hep-ph/0208251;
K.~Hagiwara, A.D.~Martin, D.~Nomura, and T.~Teubner
e-Print Archive: hep-ph/0209187.
The exact numbers in Eq. (\ref{33}) are taken from the version 1 of 
Davier et al.  In their subsequent versions, they changed it to $33.7\pm 11.2$,
which, however, has negligible effects on the results presented in this work.

\bibitem{hunter}
{\it The Higgs Hunter's Guide} by J. Gunion {\it et al.},
Addison-Wesley, New York, 1990.
\bibitem{glas}
S. Glashow and S. Weinberg, Phys. Rev. {\bf D15}, 1958 (1977).
\bibitem{cdgg}
M. Ciuchini, G. Degrassi, P. Gambino, and G.F. Giudice, Nucl. Phys. {\bf B527},
21 (1998).
\bibitem{nlo}
K. Chetyrkin, M. Misiak, and M. M\"unz, Phys. Lett. {\bf B400},
206 (1997); Erratum-{\it ibid.} {\bf B425}, 414 (1998);
A. Kagan and M. Neubert, Eur. Phys. J. {\bf C7}, 5 (1999).
\bibitem{misiak}
P. Gambino and M. Misiak, Nucl. Phys. {\bf B611}, 338 (2001).
\bibitem{darwin}
D. Chang, W. Chang, C. Chou, and W. Keung, Phys. Rev. {\bf D63}, 091301 (2001).
\bibitem{ours}
K. Cheung, C. Chou, and O.C.W.~Kong, Phys. Rev. {\bf D64}, 111301 (2001).
\bibitem{opal} 
OPAL Collaboration (prelim), P.~Ferrari {\it et.al.},
Physics Note PN475, 2001.
\bibitem{delphi} 
DELPHI Collaboration (prelim), DELPHI 2002-037-CONF-571.
\bibitem{maria-ew}
P. Chankowski, M. Krawczyk, and J. Zochowski, Eur. Phys. J. {\bf C11},
661 (1999).
\bibitem{chan}
P. Chankowski {\it et al.}, Phys. Lett. {\bf B496}, 195 (2000).
\bibitem{maria}
M. Krawczyk, hep-ph/0208076. 
\bibitem{buras}
G. Buchalla, A. Buras, and M. Lautenbacher,
Rev. Mod. Phys. {\bf 68}, 1125 (1996);
A.J.~Buras, M.~Misiak, M.~M\"unz, and S.~Pokorski,
Nucl. Phys. {\bf 424}, 374 (1994).
\bibitem{wise}
B. Grinstein, R. Springer, and M. Wise,
Nucl. Phys. {\bf B339}, 269 (1990).
\bibitem{gambino}
P. Gambino (CERN) and U. Haisch,  JHEP {\bf 0110}, 020 (2001).
\bibitem{urban}
A. Buras, A. Czarnecki, M. Misiak, and J. Urban, Nucl. Phys. {\bf B631}, 219
(2002).
\bibitem{japan}
T. Inami and C.S. Lim, Prog. Theor. Phys. {\bf 65}, 297 (1981);
Erratum, {\it ibid}, 1772.
\bibitem{ichep-b}
Plenary talk by A. Stocchi in ICHEP2002, July 2002, Amsterdam, Netherland;
talk by A.M. Eisner at the XXXVIII Rencontres de Moriond, Electroweak 
Interaction and Unified Theories, March 2003.
\bibitem{neubert}
Plenary talk by M. Neubert in SUSY02, June 2002, DESY, Hamburg, Germany.
\bibitem{abbott}
L.F. Abbott, P. Sikivie, and M. B. Wise,
Phys. Rev. D {\bf 21}, 1393 (1980);
G.G. Athanasiu, P.J. Franzini, and F.J. Gilman,
S.L. Glashow and E.E. Jenkins, Phys. Lett. {\bf 196B}, 233 (1987);
Phys. Rev. D {\bf 32}, 3010 (1985);
C.Q. Geng and J.N. Ng, Phys. Rev. D {\bf 38}, 2857 (1988).
\bibitem{pdg2002}
Particle Data Group (K. Hagiwara {\it et al.}), Phys. Rev. {\bf D66},
010001 (2002).
\bibitem{BZ}
S. Barr and A. Zee, Phys. Rev. Lett. {\bf 65}, 21 (1990); 
see also J.~Bjorken and S.~Weinberg,  Phys. Rev. Lett. {\bf 38}, 622 (1977).
\bibitem{edm}
The available literature on the class of contributions is mainly on the
closely related electric dipole moment calculations. A partial list is given by
D.~Chang,  W.-Y.~Keung, and T.C.~Yuan,  Phys. Rev. D {\bf 43}, R14 (1991);
R.G.~Leigh, S.~Paban, and R.-M.~Xu, Nucl. Phys. B {\bf 352}, 45 (1991);
C.~Kao and R.-M.~Xu, Phys. Lett. {\bf B296}, 435 (1992);
D.~Chang, W.S.~Hou, and W.-Y.~Keung,  Phys. Rev. D {\bf 48}, 217 (1993).
\bibitem{wy}
There are complicated gauge dependence notion when some of 
the class of diagrams are calculated separately. The photon Barr-Zee
diagrams with fermions running in the second loop form a gauge 
invariant set on their own. We thank Wai-Yee Keung for comments
on the issue.
\bibitem{M}
See, for example, A.~Czarnecki and W.J.~Marciano,
Phys. Rev. D {\bf 64}, 013014 (2001).
\bibitem{roco}
CDF and D0 Collaborations (M. Roco for the collaboration),
FERMILAB-CONF-00-203-E, in Proceedings of ICHEP 2000, Osaka, Japan, 
27 Jul - 2 Aug 2000.
\bibitem{lep-ew}
The LEP Collaborations Electroweak Working Group, LEPEWWG/2002-01.
\bibitem{denner}
A. Denner {\it et al.}, Z. Phys. {\bf C51}, 695 (1991).
\bibitem{pamela}
Further update or improved results may soon be available ---
P. Ferrari, private communications.
\bibitem{pamela2}
P. Ferrari, private communications.
\bibitem{langacker}
J. Erler and P. Langacker, in 
{\it Review of Particle Physics}, Phys. Rev. {\bf D66}, 010001 (2002).

\end{thebibliography}
\end{document}